\documentclass{jpp}
\usepackage{graphicx}

\usepackage[utf8]{inputenc}
\usepackage[T1]{fontenc}
\usepackage{amsmath}
\usepackage{float}
\usepackage{caption}
\usepackage{subcaption}
\usepackage{graphicx}
\usepackage{xcolor}
\usepackage{verbatim}
\usepackage{soul,xcolor}

\shorttitle{Optimization of Nonlinear Turbulence in Stellarators}
\shortauthor{P.Kim et al.}

\title{Optimization of Nonlinear Turbulence in Stellarators}

\author{P. Kim\aff{1,3}
  \corresp{\email{pk2354@princeton.edu}},
  S. Buller \aff{2}
  R. Conlin\aff{1},
  W. Dorland\aff{2, 3},
  D. W. Dudt\aff{1},
  R. Gaur\aff{1},
  R. Jorge\aff{4,5}.
  E. Kolemen\aff{1,6},
  M. Landreman\aff{2},
  N. R. Mandell\aff{6},
  \and D. Panici\aff{1}
  }

\affiliation{\aff{1}Princeton University, Princeton, New Jersey 08544, USA
\aff{2} Institute for Research in Electronics and Applied Physics, University of Maryland, College Park, MD 20742, USA
\aff{3}Department of Physics, University of Maryland, College Park, MD 20742, USA
\aff{4}Instituto de Plasmas e Fusão Nuclear, Instituto Superior Técnico, Universidade de Lisboa, 1049-001 Lisboa, Portugal
\aff{5}Department of Physics, University of Wisconsin-Madison, Madison, Wisconsin 53706, United States of America
\aff{6}Princeton Plasma Physics Laboratory, PO Box 541, Princeton NJ 08543, USA

}

\begin{document}
\setstcolor{red}
\maketitle

\begin{abstract}
We present new stellarator equilibria that have been optimized for reduced turbulent transport using nonlinear gyrokinetic simulations within the optimization loop. The optimization routine involves coupling the pseudo-spectral GPU-native gyrokinetic code {\tt GX} with the stellarator equilibrium and optimization code {\tt DESC}. Since using {\tt GX} allows for fast nonlinear simulations, we directly optimize for reduced nonlinear heat fluxes. To handle the noisy heat flux traces returned by these simulations, we employ the simultaneous perturbation stochastic approximation (SPSA) method that only uses two objective function evaluations for a simple estimate of the gradient. We show several examples that optimize for both reduced heat fluxes and good quasisymmetry as a proxy for low neoclassical transport. Finally, we run full transport simulations using the {\tt T3D} stellarator transport code to evaluate the changes in the macroscopic profiles.
\end{abstract}

\section{Introduction}
Stellarators are one of the most promising designs for magnetic confinement fusion as they are inherently steady-state and are less susceptible to current-driven instabilities \citep{Helander2014TheoryFields}. However, stellarators have historically been plagued by large collisionally-induced losses of particles and energy, associated with cross-field drifts of particles caused by the inhomogeneity and curvature of the magnetic field. In principle, stellarators can be optimized to reduce these collisional losses. In practice, impressive improvements in plasma confinement have been obtained. For example, experiments at the W7-X stellarator have shown greatly reduced neoclassical transport \citep{Beidler2021Demonstration7-X}. Furthermore, advances in stellarator optimization techniques have led to designs with precise quasisymmetry (QS) \citep{Landreman2022MagneticConfinement, Landreman2022OptimizationConfinement} and quasi-isodynamicity (QI) \citep{Goodman2022ConstructingFields}.

After neoclassical losses are minimized, confinement is limited by the anomalous transport of heat and particles by turbulence. This turbulence is driven by plasma microinstabilities on length scales of the gyroradius. For example, the ion-temperature gradient (ITG) instability is believed to be a major cause for ion-temperature clamping in W7-X, preventing heating of ions above 2 keV \citep{Beurskens2021IonPlasmas} (along with weak energy exchange between electrons and ions). While there have been recent attempts to optimize stellarators to reduce turbulence-induced transport, they have mainly relied on proxies based solely on the magnetic geometry \citep{Mynick2010OptimizingTransport, Proll2016TEMStellarators, Roberg-Clark2022ReductionOptimization} or linear simulations \citep{Jorge2023DirectDevices}. However, linear physics may not accurately predict nonlinear saturation mechanisms that ultimately determine the rate of heat and particle loss \citep{McKinney2019AQuasi-axisymmetricstellarator}. Unfortunately, using nonlinear analysis for optimization is usually very challenging. Gyrokinetics \citep{AntonsenJr.1980KineticPlasmas, Catto1978LinearizedGyro-kinetics, Frieman1982NonlinearEquilibria} is one of the most commonly-used models to study turbulence in magnetic confinement fusion devices, and is also the model used for this paper. Typical nonlinear gyrokinetics simulations usually require hundreds to thousands of CPU hours, making them infeasible to use within an optimization loop.

In this work, we demonstrate the ability to reduce turbulent losses by optimizing stellarator configurations using nonlinear turbulence simulations directly rather than  relying on proxies. In order to run nonlinear simulations inside the optimization loop, we use the new GPU-native gyrokinetic code {\tt GX} \citep{Mandell2018Laguerre-HermiteGyrokinetics,Mandell2022GX:Design}. {\tt GX} utilizes pseudo-spectral methods in velocity space. GPU acceleration combined with flexible velocity resolution allows for nonlinear {\tt GX} simulations that only take minutes to run. For this work, we focus on ITG turbulence, which contributes to major energy losses that limit plasma confinement, and so is one of the most important microinstabilities to consider for reactor design \citep{Kotschenreuther1995ComparisonInstabilities,Horton1999DriftTransport, Helander2013CollisionlessModes}. To that end, we will be running  electrostatic simulations with a Boltzmann (adiabatic) response assumed for the electrons.

The optimization is performed using the stellarator equilibrium and optimization code {\tt DESC} \citep{Dudt2020DESC:Solver,Panici2023TheComputations,Conlin2023TheMethods,Dudt2023TheOptimization}. {\tt DESC} also uses pseudo-spectral methods and directly solves the ideal MHD force-balance equation to compute the magnetic equilibrium. The quantities computed from the resulting magnetic fields are used as input for {\tt GX}. Stochastic optimization methods are used to robustly handle the noisy landscape of turbulent heat fluxes.

The paper is organized as follows. The stochastic optimization method used in this work is described in Section \ref{Optimization Methods}. Results are shown in Section \ref{Results}, with analysis on the effects of global magnetic shear for reduced turbulence explored in Section \ref{Mechanisms for Reduced Turbulence}. Transport simulations using the {\tt T3D} stellarator transport code \citep{Qian2022StellaratorGX} are shown in Section \ref{trinity sim}. Finally, the conclusions follow in Section \ref{Conclusions}. 

\section{Optimization Methods}\label{Optimization Methods}
In this optimization routine, we seek to minimize the nonlinear heat flux returned by {\tt GX}. Specifically, {\tt GX} returns as output the time-trace of the heat flux (normalized to gyro-Bohm units). An example of some heat traces is shown in Figure \ref{heat flux trace}. At the beginning of the simulation, linear growth of the fastest-growing instability dominates. However, eventually nonlinear effects cause the heat flux to decrease and saturate to a statistical steady-state. We use the time-average of this steady-state flux as our heat flux objective $f_Q$. To take the time-average, we take the second half of the time-trace and compute the weighted Birkoff average
\begin{equation}
    f_Q = \frac{1}{I}\sum_i^N e^{-i/\left(N\left(1-i/N\right)\right)}q_i  
\end{equation}
where the sum is over each point in the trace (with $N$ total points) and $I = \sum_i^N e^{-i/\left(N\left(1-i/N\right)\right)}$ is a normalization factor. This gives greater weight to values in the middle of the trace.

Since {\tt GX} is a local flux-tube gyrokinetic code, each simulation is on a single field line on a single surface specified by the field line label $\alpha$ and the radial coordinate $\rho = \sqrt{\psi/\psi_b}$ (where $B = \nabla \psi \times \nabla \alpha$ and $\psi$ is the toroidal magnetic flux). Therefore, $f_Q$ is also computed for only a single field line and surface. For all of the optimization examples, we only simulate on the $\rho = \sqrt{\psi/\psi_b} = \sqrt{0.5}$ surface. and the $\alpha = 0$ (except for the multiple field line examples in Section \ref{multiple alpha case}) field line. However, we will run post-processing simulations across different surfaces and field lines.

To run these simulations, {\tt GX} requires a set of geometric quantities that can be computed from numerical equilibria. Given that the magnetic field is written in Clebsch form $\mathbf{B} = \nabla \psi \times \nabla \alpha$, the set of quantities needed are
    \begin{align}\label{geo quantitities}
    \begin{split}
    \mathcal{G} &= \{B, \mathbf{b \cdot \nabla z}, |\nabla \psi|^2, |\nabla \alpha|^2, \nabla \psi \cdot \nabla \alpha, \\ &(\mathbf{B} \times \nabla B) \cdot \nabla \psi, (\mathbf{B} \times \nabla B) \cdot \nabla \alpha, (\mathbf{b} \times \mathbf{\kappa}) \cdot \nabla \alpha\},
    \end{split}
\end{align}
where $\mathbf{b} = \mathbf{B}/B$, $\alpha$ is the straight field line label, $\kappa = \mathbf{b} \cdot \nabla \mathbf{b}$ is the curvature, and $z$ is some coordinate representing the distance along the field line. For these simulations, the geometric toroidal angle $\phi$ is used for $z$. All of these quantities are easily computed using utility functions in {\tt DESC}. 

\begin{figure}
    \centering
    \includegraphics[scale=0.5]{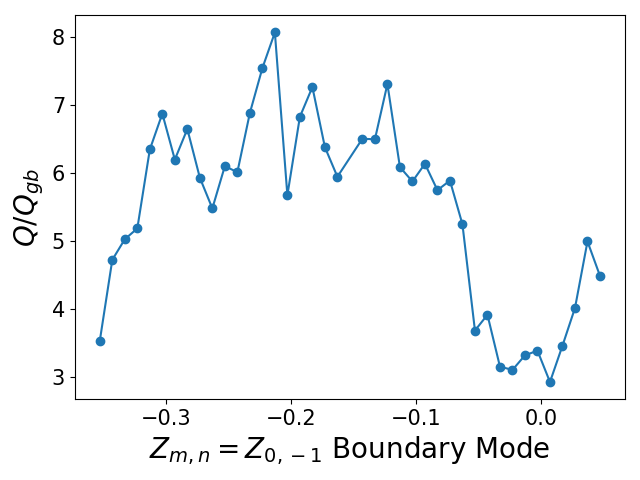}
    \caption{The time-averaged nonlinear heat flux computed by {\tt GX} when scanning across the $Z_{0,-1}$ boundary mode.}
    \label{fig:boundary scan}
\end{figure}

One issue with directly minimizing nonlinear heat fluxes is that their time-traces are often very noisy. Even the resulting time averages are usually very noisy in parameter space. This can be seen in Fig. \ref{fig:boundary scan} showing the time-averaged nonlinear heat flux when scanning over the $Z_{m,n} = Z_{0,-1}$ boundary mode of the initial equilibrium used in this study. Optimizers designed for smooth objectives may easily get stuck in local minima and make very little progress.
To help with this issue, we use the simultaneous perturbation stochastic approximation (SPSA) method \citep{Spall1987AEstimates} to minimize the heat fluxes. In this algorithm, the $i^{th}$ component of the gradient at point $\mathbf{x_n}$ is approximated as
\begin{equation}
    \mathbf{\hat{g}_n\left(x_n\right)}_i = \frac{f\left(\mathbf{x_n} + k\mathbf{c_n}\right) - f\left(\mathbf{x_n} - k\mathbf{c_n}\right)}{2kc_{ni}},
\end{equation}
where $\mathbf{c_n}$ is a random perturbation vector whose components are sampled from a Rademacher distribution, and $k$ is a finite-difference step-size. Therefore, we are effectively using the finite difference method in all directions at once. Finally, for this method, rather than specifying stopping tolerances, we instead specify a maximum number of iterations.

The main feature of the SPSA method is that it only requires two objective function measurements per iteration. This makes it suitable for high-dimensional optimization problems whose objective functions are expensive to compute. SPSA can also robustly handle noisy objective functions, and so has been used for simulation optimization, including Monte-Carlo simulations \citep{Chan2003OptimisationApproximation}.

In order to still use automatic differentiation and smooth optimization methods for other objectives like quasisymmetry residuals, we split the optimization routine into two parts. We first minimize the turbulent heat flux using stochastic gradient descent. Next, we minimize the quasisymmetry residuals using a least-squares method that utilizes automatic differentiation. This order is chosen arbitrarily, but the reverse ordering can also work. The two objectives are thus
\begin{align}
    f_1 &= f_{Q}^2 + \left(A - A_{target}\right)^2 \\
    f_2 &= f_{QS}^2 + \left(A - A_{target}\right)^2,
\end{align}
where $f_Q$ is the heat flux from {\tt GX}, and $A$ is the aspect ratio. For these optimizations we target an aspect ratio of $A_{target} = 8$. We use the two-term quasisymmetry objective, where a magnetic field is quasisymmetric if the quantity
\begin{equation}
    C = \frac{\left(\mathbf{B} \times \nabla \psi\right) \cdot \nabla B}{\mathbf{B} \cdot \nabla B}
\end{equation}
is a flux function. Computationally, we evaluate the equivalent form
\begin{equation}
    f_{QS} = \left(M - \iota N\right)\left(\mathbf{B} \times \nabla \psi\right)\cdot \nabla B - \left(MG + NI\right)\mathbf{B} \cdot \nabla B,
\end{equation}
at several different flux surfaces and try to minimize the resulting values. For this study, we always choose flux surfaces at $\rho = 0.6, \ 0.8, \text{and} \ 1$. We target quasi-helical (QH) symmetry, so that the helicity is $(M, N) = (1, 4)$.

The {\tt GX} simulation parameters used in the optimization loop and for post-processing, along with their justification, are in Appendix \ref{app sim param}.

Finally, the optimization routine is performed in stages. Each stage increments the maximum boundary Fourier mode being optimized over. For example in the first stage only boundary modes with modes $m,n$ satisfying $|m| \leq 1$ and $|n| \leq 1$ are used as optimization variables. In the next stage boundary modes with $|m| \leq 2$ and $|n| \leq 2$ are used. The $m=0,n=0$ mode is excluded to prevent the major radius from changing. This reduces the number of optimization variables at the beginning of the optimization and "warm-starts" each successive stage. This type of method has been used with great success for previous stellarator optimization results, including for optimizing quasisymmetry \citep{Landreman2022MagneticConfinement}. After each stage, we start the next stage with an equilibrium chosen from the previous stage that had achieved both low heat flux and low quasisymmetry error. No strict rule is used, and the choice depended on several factors like the value of the gradient estimate, and the $\iota$ profile.

\section{Results}\label{Results}
\subsection{Turbulence Optimization}\label{turbulence only opt}
As an initial test of our stochastic optimization method, we begin by solely optimizing for turbulence. The optimization routine is the same as described in the previous section, except instead of optimizing for both quasisymmetry and aspect ratio in the second half of each iteration, we only optimize for the target aspect ratio. Usually, combining the turbulence and aspect ratio objectives as is done in the first part is insufficient to reach our target aspect ratio. We have to force the optimizer to take more aggressive steps due to the simulation noise, which makes it more difficult to maintain the desired aspect ratio. For this optimization (and all of the examples in this work), the initial equilibrium is an approximately quasi-helically symmetric equilibrium with an aspect ratio of 8 and 4 field-periods.

\begin{figure}
    \centering
    \includegraphics[scale=0.5]{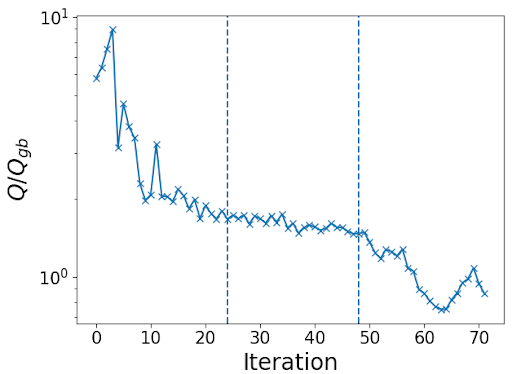}
    \caption{The normalized heat flux across each iteration. The dashed lines represent an increase in the maximum boundary mode number being optimized.}
    \label{fig:qflux iteration}
\end{figure}

Figure \ref{fig:qflux iteration} shows the normalized nonlinear heat flux across each iteration of the optimization process. Due to the stochastic nature of optimization routine, the initial gradient estimates are very poor, leading to an increase in the heat flux in the first several iterations. However, as the optimization continued, there is eventually a rapid decrease in the heat flux as the gradient approximation begins to track the true gradient. The optimizer continues to steadily decrease the heat flux for most of the remaining iterations. Note that since we specify a maximum number of iterations rather than stopping tolerances, the optimization ends even when the heat flux had increased slightly at the end.

\begin{figure}
     \centering
     \begin{subfigure}[c]{0.45\textwidth}
         \includegraphics[width=\textwidth]{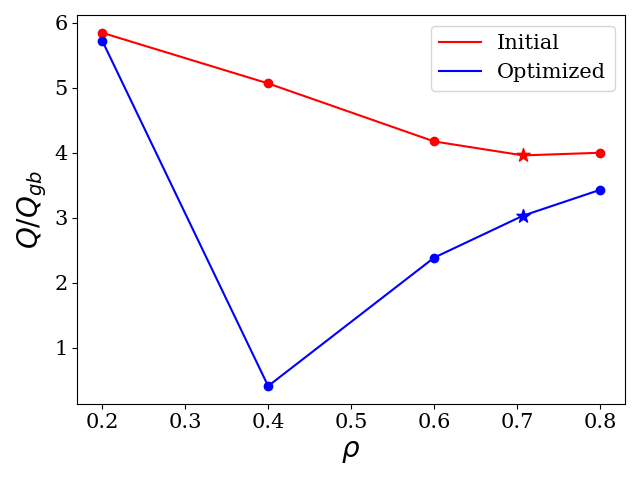}
         \caption{}
         \label{heat flux trace}
     \end{subfigure}
     \hfill
     \begin{subfigure}[c]{0.45\textwidth}
         \includegraphics[width=\textwidth]{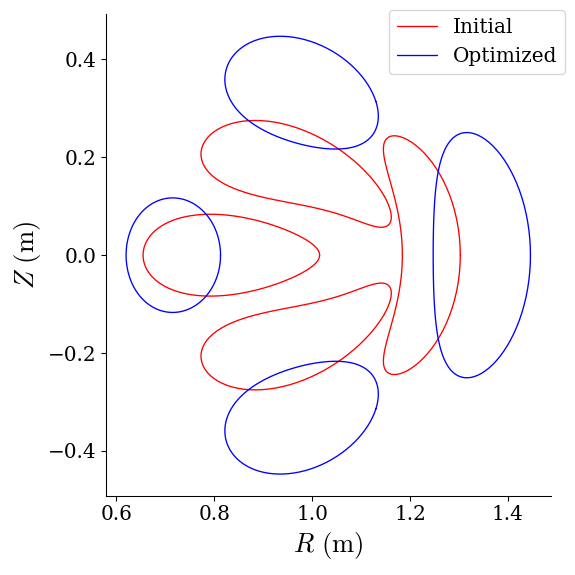}
         \caption{}
         \label{cross sections}
     \end{subfigure}
     \caption{(a) A scan of the nonlinear heat flux for the initial (red) and optimized (blue) equilibria across different flux surfaces. (b) The cross-sections of the boundary magnetic flux surface of the initial (red) and optimized (blue) configurations. The star in (a) indicates the $\rho = \sqrt{0.5}$ surface that was chosen for the optimization loop.}
     \label{turbulence only}
\end{figure}

A scan of the nonlinear heat fluxes across $\rho$ and optimized cross-sections of the flux surfaces are shown in Fig. \ref{turbulence only}. For this simulation (and all of the simulations in this section), the resolution is increased to the values in Table \ref{sim table post}. As seen in those plots, some surfaces see moderate to drastic improvements to the nonlinear heat flux. In particular, the $\rho = 0.4$ surfaces has a reduction of about an order of magnitude. However, other surfaces see much less improvement, such as at $\rho = 0.2$ and $\rho = 0.8$. It is not completely well-understood why that is, but it could be related to the fact that only the $\rho = \sqrt{0.5}$ surface is being optimized. This does, reveal a potential weakness in the optimization strategy that will be addressed in future work. Nevertheless, the following sections show much more uniform improvement. Interestingly, based off of the surface boundary plots, the cross-sections of the optimized equilibria seem much less strongly-shaped than in the initial equilibrium. Instead, the magnetic axis has significantly more torsion.

\begin{figure}
    \centering
    \includegraphics[scale=0.4]{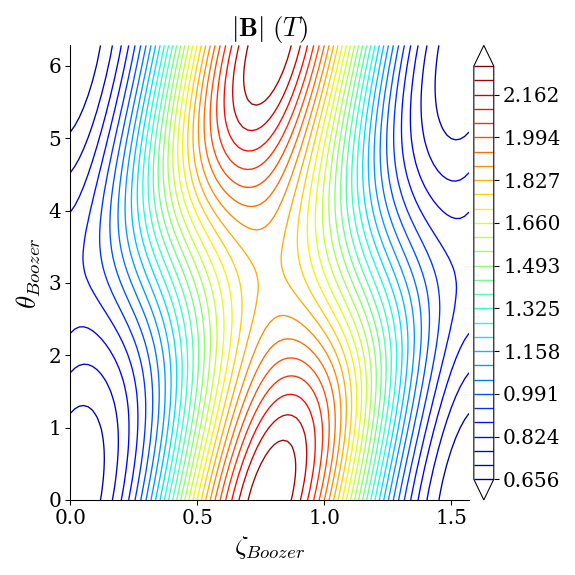}
    \caption{The $|\mathbf{B}|$ contours in Boozer coordinates, which resemble contours of a QI equilibrium.}
    \label{fig:|B| turbulence only}
\end{figure}

The contours of $|\mathbf{B}|$ in Boozer coordinates are plotted in Fig. \ref{fig:|B| turbulence only}. Surprisingly, the contours seem to resemble those from quasi-isodynamic equilibria despite not including a QI term in the objective functions. This may be related to the fact that it seems like the optimizer favored a high mirror ratio. Additionally, it is well-known that QI stellarators can be optimized to have the maximum-J property, which has numerous benefits including improved confinement of fast ions, neoclassical confinement of thermal ions \citep{Sanchez2023A,Velasco2023RobustFields} and enhanced stabilization against trapped-electron modes (TEM) \citep{Proll2012ResilienceInstabilities,Helander2013CollisionlessModes, Helander2014TheoryFields}. It's been further theorized and shown in gyrokinetic simulations of W7-X that possessing the maximum-J property can also be beneficial for ITG turbulence as well \citep{Proll2022TurbulenceGradient}. However, the TEM studies required modeling the full kinetic electron dynamics. In this study, we only use a Boltzmann response for the electrons.

\subsection{Combined Turbulence-Quasisymmetry Optimization}\label{turbulence-qs}
\begin{figure}
     \centering
     \begin{subfigure}[c]{0.45\textwidth}
         \centering
         \includegraphics[width=\textwidth]{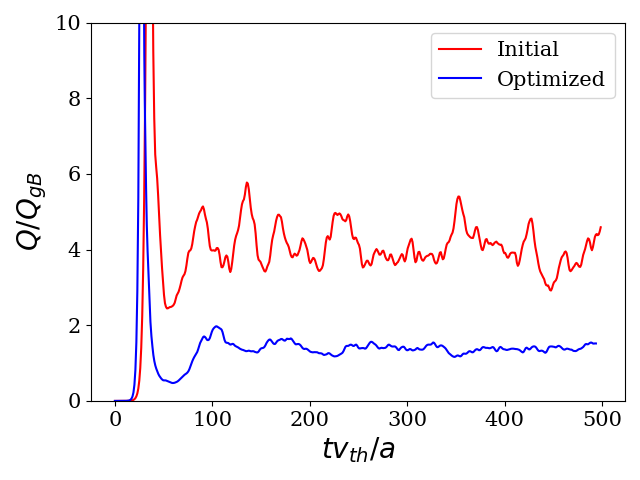}
         \caption{}
         \label{heat flux trace}
     \end{subfigure}
     \hfill
     \begin{subfigure}[c]{0.45\textwidth}
         \centering
         \includegraphics[width=\textwidth]{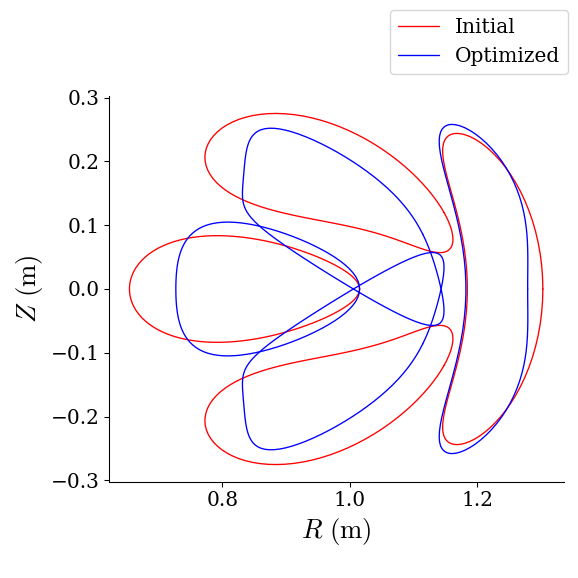}
         \caption{}
         \label{cross sections}
     \end{subfigure}
     \caption{(a) The time-traces of the nonlinear heat fluxes of the initial (red) and optimized (blue) configurations. (b) The cross-sections of the magnetic flux surfaces of the initial (red) and optimized (blue) configurations.}
     \label{turbulence-qs trace and boundaries}
\end{figure}

Next, we include the two-term quasisymmetry objective in the second part of the optimization loop. The final heat flux traces and optimized cross-sections of the flux surfaces are shown in Fig. \ref{turbulence-qs trace and boundaries}.  The cross sections in Fig. \ref{cross sections} show relatively modest changes in the shape of optimized stellarator. However, the time-trace (simulated at the $\psi/\psi_b = 0.5$ surface) in Fig. \ref{heat flux trace} shows about a factor of 3 decrease in the nonlinear heat flux.

\begin{figure}
     \centering
     \begin{subfigure}[c]{0.45\textwidth}
         \centering
         \includegraphics[width=\textwidth]{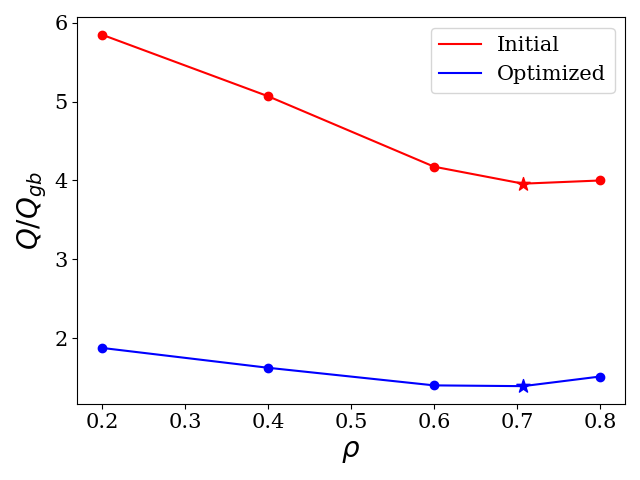}
         \caption{}
         \label{heat flux rho scan}
     \end{subfigure}
     \hfill
     \begin{subfigure}[c]{0.45\textwidth}
         \centering
         \includegraphics[width=\textwidth]{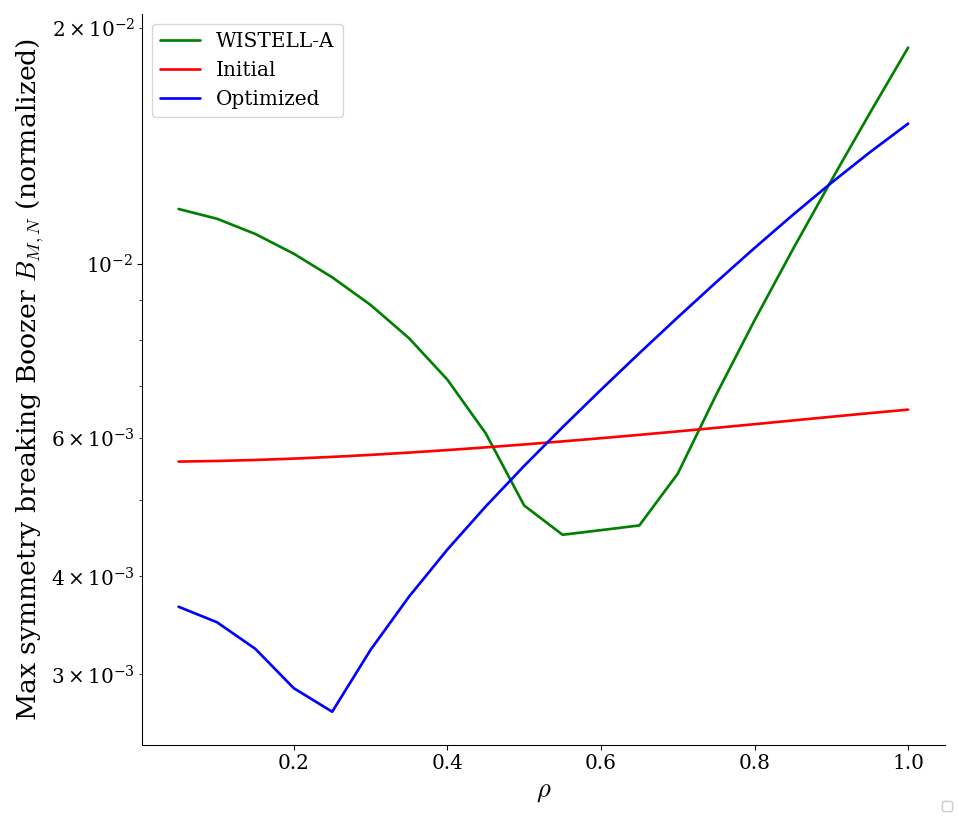}
         \caption{}
         \label{qs rho scan}
     \end{subfigure}
     \caption{Scans of the nonlinear heat flux (a) and maximum symmetry breaking modes (b) across different radial locations. The star in (a) indicates the $\rho = \sqrt{0.5}$ surface that was chosen for the optimization loop.}
     \label{turbulence qs rho scan}
\end{figure}

\begin{figure}
     \centering
     \begin{subfigure}[b]{0.45\textwidth}
         \centering
         \includegraphics[width=\textwidth]{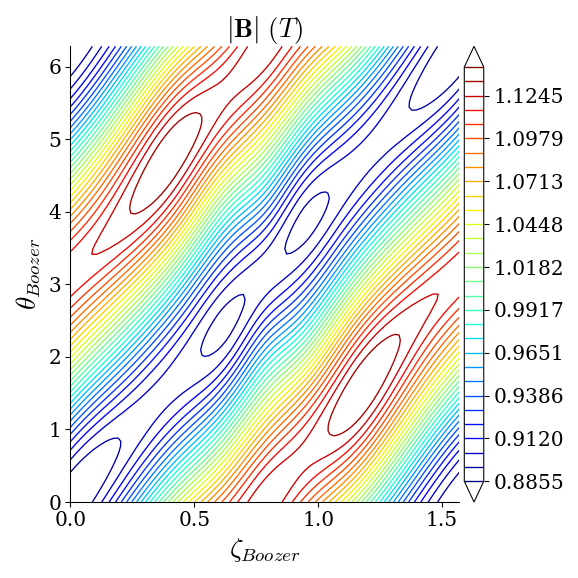}
         \caption{}
         \label{|B| contours initial}
     \end{subfigure}
     \hfill
     \begin{subfigure}[b]{0.45\textwidth}
         \centering
         \includegraphics[width=\textwidth]{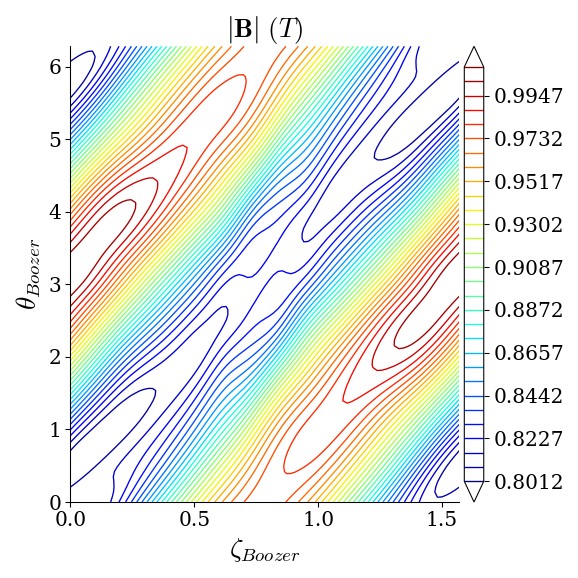}
         \caption{}
         \label{|B| contours final}
     \end{subfigure}
     \caption{The $|\mathbf{B}|$ contours in Boozer coordinates for the initial (a) and final (b) equilibria.}
     \label{|B| plots}
\end{figure}

To ensure that the nonlinear heat flux was reduced throughout the plasma volume, we again ran simulations at radial locations of $\rho = 0.2$, $0.4$, $0.6$, and $0.8$. The resulting heat fluxes as well as the maximum symmetry breaking residuals across $\rho$ are plotted in Fig. \ref{turbulence qs rho scan}. As seen in the plots, despite optimizing only for the $\rho = \sqrt{0.5}$ surface, the heat flux is reduced throughout the plasma volume. With regard to quasisymmetry, for $\rho < 0.5$, the optimized equilibrium has smaller maximum symmetry-breaking modes than the initial equilibrium. However, this reverses in the outer surfaces. This is unexpected considering that the quasisymmetry residuals were computed at $\rho = 0.6, \ 0.8 \ $ and $1$. Nevertheless, the degree of symmetry-breaking is still comparable to WISTELL-A \citep{Bader2020AdvancingStellarators}, another optimized QH stellarator. Indeed, the $|\mathbf{B}|$ plots in Fig. \ref{|B| plots} show contours characteristic of a quasi-helically symmetric stellarator (plotted at the $\psi = 0.5$ surface).

\begin{figure}
     \centering
     \begin{subfigure}[b]{0.45\textwidth}
         \centering
         \includegraphics[width=\textwidth]{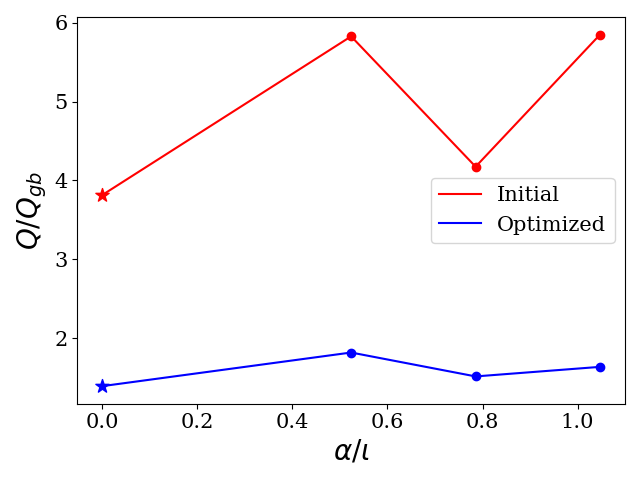}
         \caption{}
         \label{turbulence qs alpha scan}
     \end{subfigure}
     \hfill
     \begin{subfigure}[b]{0.45\textwidth}
         \centering
         \includegraphics[width=\textwidth]{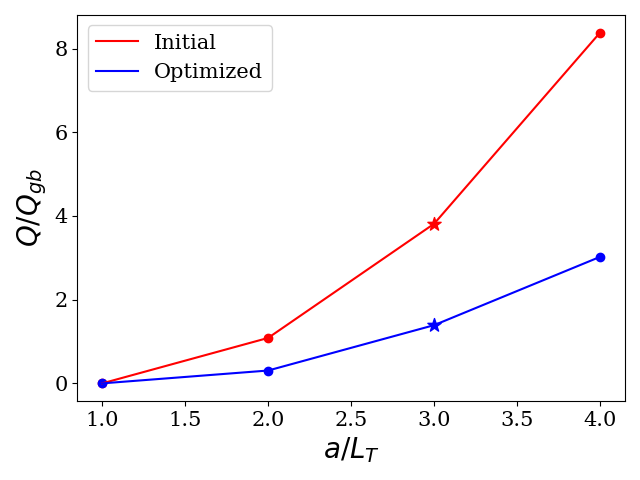}
         \caption{}
         \label{a/Lt scan}
     \end{subfigure}
     \caption{The heat fluxes across different field lines $\alpha$ (a) and $a/L_T$ (b) for the initial (red) and optimized (blue) configurations. The star indicates the $\alpha = 0$ field line and the $a/L_T = 3$ temperature gradient that were chosen for the optimization loop.}
\end{figure}

Unlike in tokamaks, different field lines in stellarators experience different curvatures, magnetic fields, etc. and so may have very different fluxes \citep{Dewar1983BallooningSystems, Faber2015GyrokineticStellarator}. Therefore, we also run simulations across $\alpha$ at the radial location $\rho = \sqrt{0.5}$. The resulting scan is shown in Fig. \ref{turbulence qs alpha scan}. At each $\alpha$ simulated, the optimized stellarator has a lower heat flux, indicating that the heat flux has been reduced across the entire flux surface.

\begin{figure}
    \centering
    \includegraphics[scale=0.75]{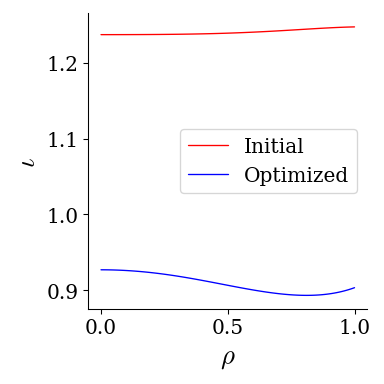}
    \caption{The $\iota$ profiles for the initial (red) and optimized (blue) equilibria.}
    \label{fig:iota scan}
\end{figure}

It is interesting to see that, for the initial geometry, there are large variations in the heat flux. This might indicate that some flux tubes are too short to sample enough area on the flux surface or that there is some coupling between the flux tubes. However, it should be noted that the $\iota$ of this equilibrium (plotted in Fig. \ref{fig:iota scan}) is close to $5/4$. Recent work has shown a very large variation across $\alpha$ on low-order rational surfaces \citep{Buller2023LinearStellarators}. In comparison, the optimized equilibrium has an $\iota$ that is slightly above 0.9. This could have negative impacts on the optimization routine as the optimizer may choose to approach equilibria with $\iota$ near low-order rationals. This could then lead to not only misleading heat fluxes, but also poor MHD stability. More work is needed to investigate the limitations of flux-tube codes, the effects of rational surfaces, and the resulting consequences for optimization.

In a fusion reactor, since the transport is very stiff, the temperature profiles may evolve to approach the critical temperature gradient (the temperature gradient at which the heat flux is zero). Therefore, a decrease in the heat flux is less useful if the critical temperature gradient also decreases significantly. To verify that the critical temperature gradient did not decrease, we also ran several nonlinear simulations with different temperature gradients at $\rho = \sqrt{0.5}$. The results are shown in Fig. \ref{a/Lt scan} which shows that the critical temperature gradient does not seem to change significantly. It would be more beneficial if the critical temperature gradient had also increased. Specifically targeting the critical temperature gradient will be the focus of future work.

We also run linear simulations for the initial and optimized equilibria, with the growth rates shown in Fig. \ref{linear kx0} at $k_x = 0$ and Fig. \ref{linear kx0.4} at $k_x = 0.4$. Although the optimized equilibria have lower heat fluxes, they have a significantly higher peak growth rate at $k_x = 0$. On the other hand, it has lower growth rates at lower $k_y$, and one would expect that the nonlinear heat flux scales like $\gamma/\langle k_{\perp}^2\rangle$ (where the angle brackets indicate a flux-surface average) \citep{Mariani2018IdentifyingFluxes}. Therefore, this might result in the observed lower nonlinear heat fluxes. However, this trend reverses for the simulations at $k_x = 0$. Furthermore, recent work has shown that both peak growth rates and the quasilinear $\gamma/\langle k_{\perp}^2\rangle$ estimate are both poor proxies for the nonlinear heat flux \citep{Buller2023LinearStellarators}.

\begin{figure}
     \centering
     \begin{subfigure}[c]{0.45\textwidth}
         \centering    \includegraphics[width=\textwidth]{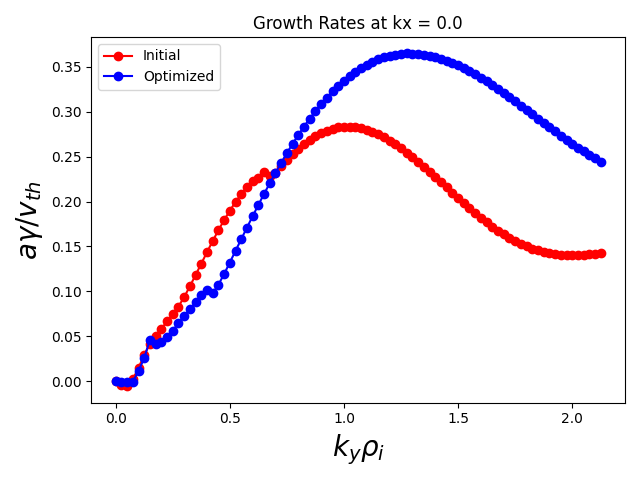}
         \caption{}
         \label{linear kx0}
     \end{subfigure}
     \hfill
     \begin{subfigure}[c]{0.45\textwidth}
         \centering       \includegraphics[width=\textwidth]{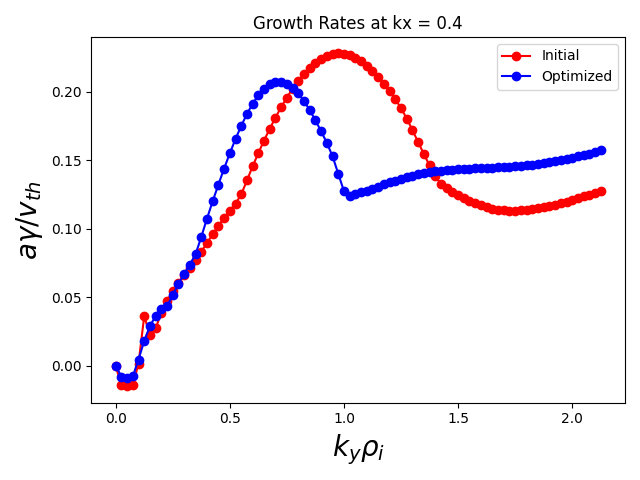}
         \caption{}
         \label{linear kx0.4}
     \end{subfigure}
     \caption{The linear growth rates across different $k_y$ at $k_x = 0$ (a) and $k_x = 0.4$ (b)}
     \label{linear sim}
\end{figure}

\begin{figure}
     \centering
     \begin{subfigure}[c]{0.45\textwidth}
         \centering    \includegraphics[width=\textwidth]{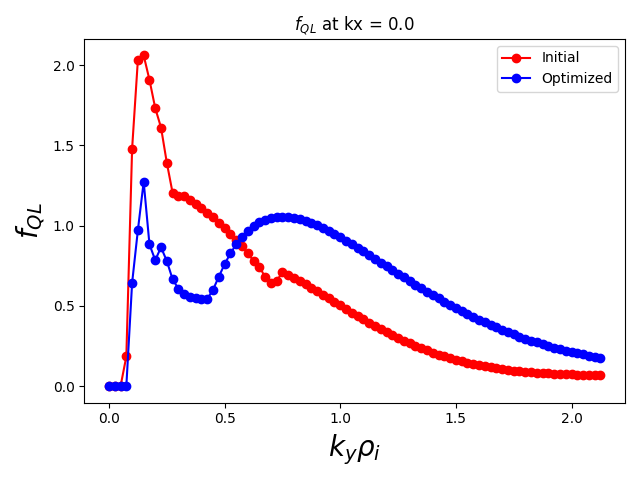}
         \caption{}
         \label{linear kx0}
     \end{subfigure}
     \hfill
     \begin{subfigure}[c]{0.45\textwidth}
         \centering       \includegraphics[width=\textwidth]{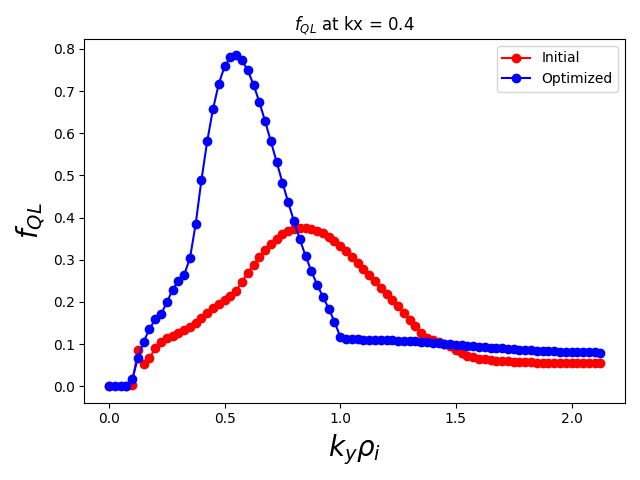}
         \caption{}
         \label{linear kx0.4}
     \end{subfigure}
     \caption{The quasilinear estimate for the initial (red) and optimized (blue) configurations at $k_x = 0$ (left) and $k_x = 0.4$ (right).}
     \label{ql sim}
\end{figure}

As a preliminary example, we replicate the plots from Fig. \ref{linear sim} but with a quasi-linear estimate of the form
\begin{equation}
    f_{QL}(k_y) = \frac{\gamma(k_y)}{\langle k_{\perp}^2 \rangle}
\end{equation}
where $\langle k_{\perp}^2 \rangle$ is
\begin{equation}
    \langle k_{\perp}^2 \rangle = \frac{\int dz k_{\perp}^2 \phi^2 \sqrt{g}}{\int dz \phi^2 \sqrt{g}}
\end{equation}
Here, $\phi$ is the electrostatic potential, $\sqrt{g}$ is the Jacobian, and the integration is along the simulated field line. This expression is similar to the one used in \cite{Mariani2018IdentifyingFluxes}. The resulting plots are shown in Fig. \ref{ql sim}. We have enforced that $\min(f_{QL}) = 0$, which assumes that only growing modes contribute to the flux. As seen in the plots, while the optimized configuration has a smaller $f_{QL}$ at smaller $k_y$ at $k_x = 0$, it has a much larger $f_{QL}$ at $k_x = 0.4$. This makes it challenging to use such a quasilinear estimate of the heat flux as a proxy, necessitating our approach of directly computing the nonlinear heat flux.

\begin{figure}
    \centering
    \includegraphics[scale=0.5]{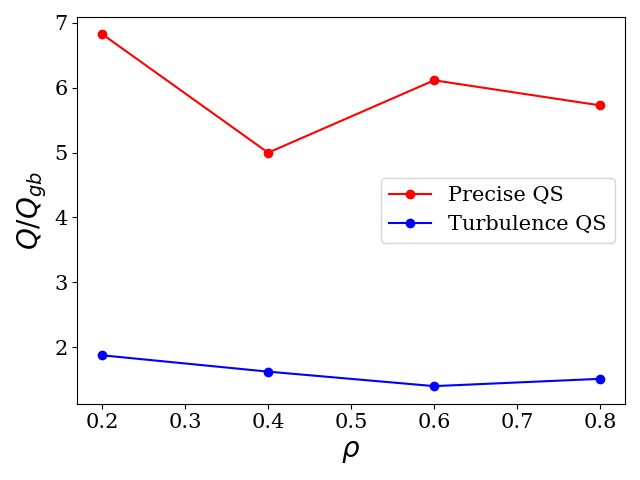}
    \caption{The heat fluxes across $\rho$ for a precise QH equilibrium (red) and the turbulence optimized equilibrium (blue).}
    \label{qflux precise}
\end{figure}

Finally, we compare the optimized stellarator to another configuration solely optimized for quasisymmetry. The plot of the heat fluxes across different surfaces is shown in Fig \ref{qflux precise}. The heat fluxes for the precise QH equilibrium are higher than those from the approximately QH equilibrium used as the initial point. This shows that just optimizing for quasisymmetry can be detrimental for turbulence, and stresses the need to optimize for both.

\subsection{Optimization with Multiple field lines}\label{multiple alpha case}
As described in the previous section, several physical and computational factors can lead to large variations in the heat fluxes at different $\alpha$. To avoid this issue, we rerun the optimization loop but simulate on field lines of both $\alpha = 0$ and $\alpha = \pi \iota/4$. This makes the new objective function

\begin{equation}
    f_Q = f_{Q,\alpha=0}^2 + f_{Q,\alpha=\pi \iota/4}^2 + (A - A_{target})^2.
\end{equation}

This new objective function serves both as a way of reducing the heat flux across multiple field lines while also testing our stochastic optimizer against additional objectives. Similar situations include trying to optimize across different flux surfaces, different temperature/density gradients, etc.

\begin{figure}
     \centering
     \begin{subfigure}[b]{0.45\textwidth}
         \centering
         \includegraphics[width=\textwidth]{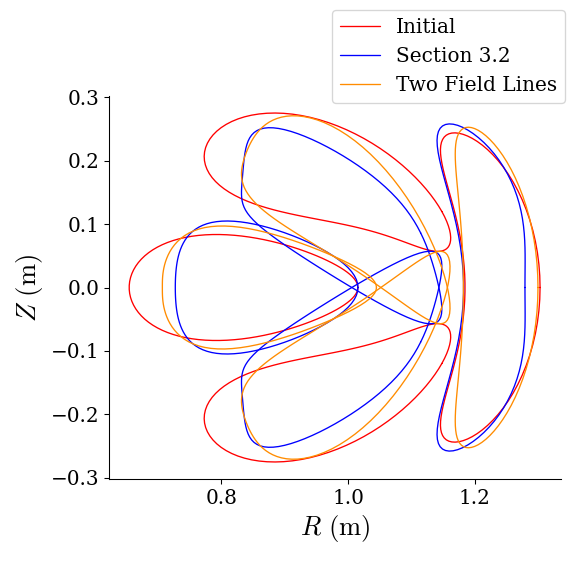}
         \caption{}
         \label{}
     \end{subfigure}
     \hfill
     \begin{subfigure}[b]{0.45\textwidth}
         \centering
         \includegraphics[width=\textwidth]{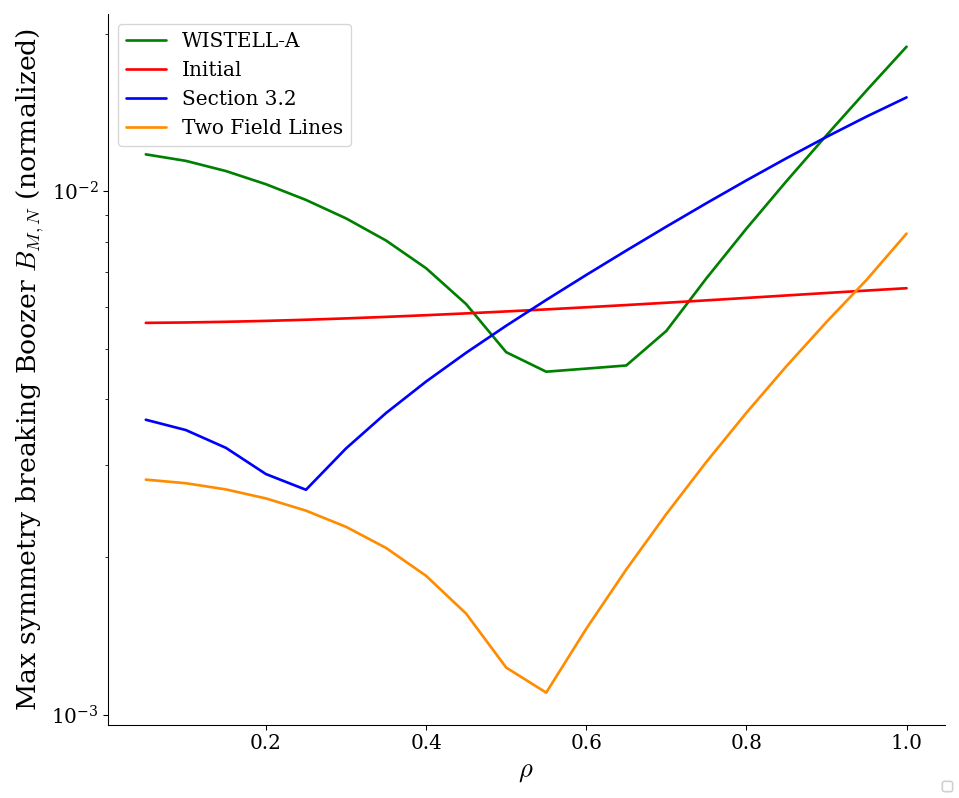}
         \caption{}
         \label{}
     \end{subfigure}
     \begin{subfigure}[b]{0.45\textwidth}
         \centering
         \includegraphics[width=\textwidth]{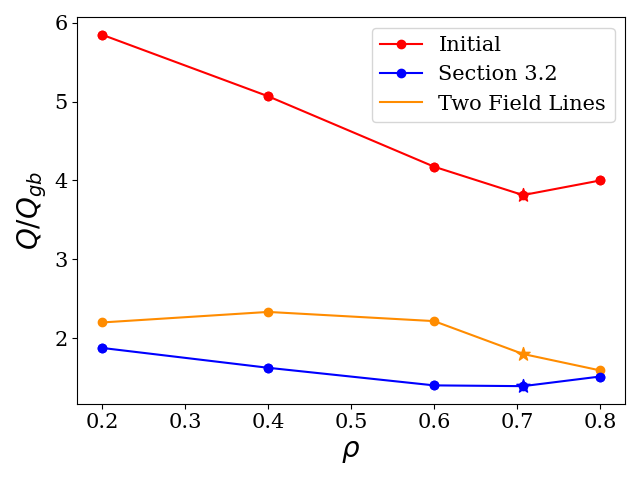}
         \caption{}
         \label{}
     \end{subfigure}
     \begin{subfigure}[b]{0.45\textwidth}
         \centering
         \includegraphics[width=\textwidth]{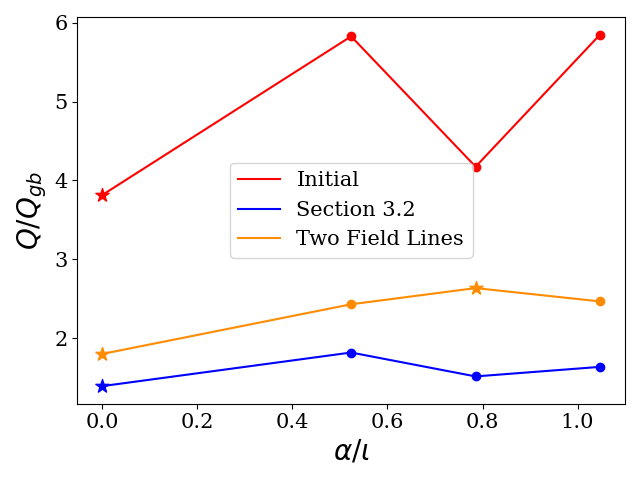}
         \caption{}
         \label{alpha scan}
     \end{subfigure}
     \caption{(a) The cross sections for the initial equilibrium (red), the single field line optimized equilibrium (blue)from Section \ref{turbulence-qs}, and the new two field line optimized equilibrium. (b) The maximum symmetry-breaking modes for all three equilibria and WISTELL-A. (c) The heat flux scans across $\rho$. (d) The heat flux scans across $\alpha$. The stars in (c) and (d) indicate parameters that were chosen for optimization.}
     \label{multiple alpha}
\end{figure}

The plots in Fig. \ref{multiple alpha} show the cross-sections, maximum symmetry breaking modes, and heat fluxes across $\rho$ and $\alpha$ for the initial equilibrium and the final optimized equilibria after optimizing at one and two field lines. While this new equilibrium has higher heat fluxes than when optimizing for just a single field line, it instead has a lower maximum symmetry breaking modes and better quasisymmetry.

The scans across multiple $\alpha$ in Fig. \ref{alpha scan} show about a 50$\%$ variation in the heat flux compared to the $\alpha = 0$ point. Unfortunately, this is larger than the approximately 30$\%$ variation in the single-field line case, and comparable to the relative variation in the initial equilibrium. Nevertheless, the heat fluxes across each $\alpha$ are still 2-3x smaller than in the initial equilibrium and this shows that the stochastic optimizer is still effective with additional terms in the objective function. More investigation is needed to more effectively and precisely optimize with multiple field lines.

\subsection{Fluid Approximation}\label{fluid opt}
In this final case, we change the simulation parameters within the optimization loop to approach the fluid limit in {\tt GX}. This is based off of previous work that showed a fluid approximation with only a few velocity moments can accurately match the true heat fluxes at higher velocity resolution at moderate collision frequencies \citep{Buck2022AGX}. That study was motivated by a recently developed three-field model for the density, temperature, and momentum to approximate growth rates and nonlinear heat fluxes \citep{Hegna2018TheoryTransport}.

To approximate this model, we reduce the velocity resolution to 4 Hermite moments and 2 Laguerre moments, as in the gyrofluid model. While normally this requires closure relations like in \citep{Beer1995GyrofluidTokamaks,Mandell2018Laguerre-HermiteGyrokinetics}, we instead increase the collisionality to damp the high wavenumber modes and increase the temperature gradient to $a/L_T = 5$. We employ a Dougherty collision operator \citep{Dougherty1964ModelSolution}.

\begin{figure}
     \centering
     \begin{subfigure}[c]{0.45\textwidth}
         \centering
         \includegraphics[width=\textwidth]{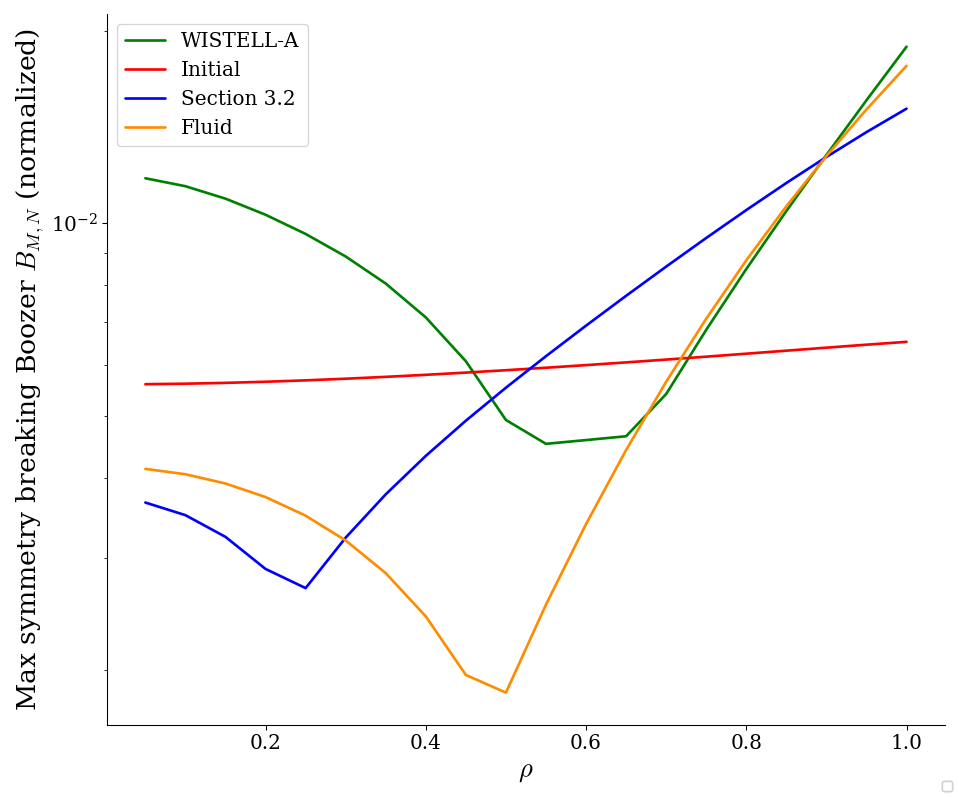}
         \caption{}
         \label{}
     \end{subfigure}
     \hfill
     \begin{subfigure}[c]{0.45\textwidth}
         \centering
         \includegraphics[width=\textwidth]{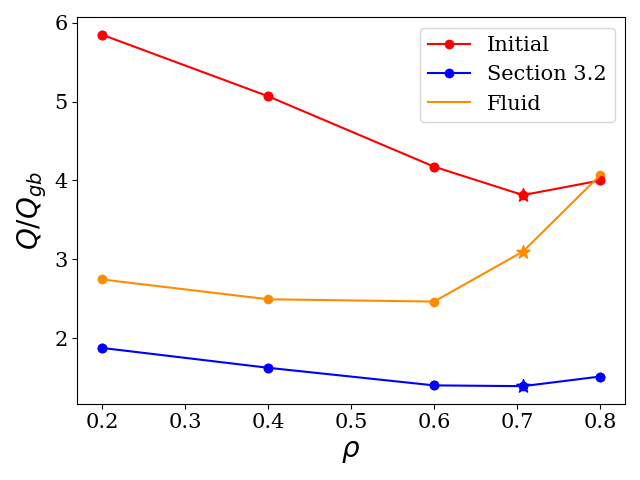}
         \caption{}
         \label{alpha scan psi = 0.5}
     \end{subfigure}
     \caption{The maximum symmetry-breaking modes (a) and the heat flux scan across $\rho$ (b) for the initial equilibrium, the kinetic case from Section \ref{turbulence-qs}, and the fluid case (as well as WISTELL-A for the left plot). The star in (b) indicates the $\rho = \sqrt{0.5}$ surface that was chosen for the optimization loop.}
     \label{fluid}
\end{figure}

The plots of the maximum-symmetry breaking modes and the heat flux across $\rho$ for the kinetic and fluid cases are shown in Fig. \ref{fluid}. While the new equilibrium does not achieve lower heat fluxes than in the kinetic case, it still achieves about a factor of 2 reduction compared to the initial equilibrium except at $\rho = 0.8$ despite using very different physical parameters. The fluid case retains comparable levels of quasisymmetry as well. These results indicate that the optimization routine is robust against varying levels of fidelity. This opens new possibilities of potentially varying the fidelity across iterations. This can either decrease the computational cost further or allow us to increase other resolution parameters.

\section{Effects of Magnetic Shear on Reduced Turbulence}\label{Mechanisms for Reduced Turbulence}

It is well-known since \citet{Greene1981TheModes} that sufficient positive global magnetic shear can be destabilizing for intermediate to high wavenumber instabilities like the ballooning mode. An analogous relationship for transport was investigated in \citet{Kotschenreuther1995QuantitativeEffects}, where it was found that while increased shear can reduce thermal transport, it can also negatively impact the critical gradient.

\begin{figure}
    \centering
    \includegraphics[scale=0.5]{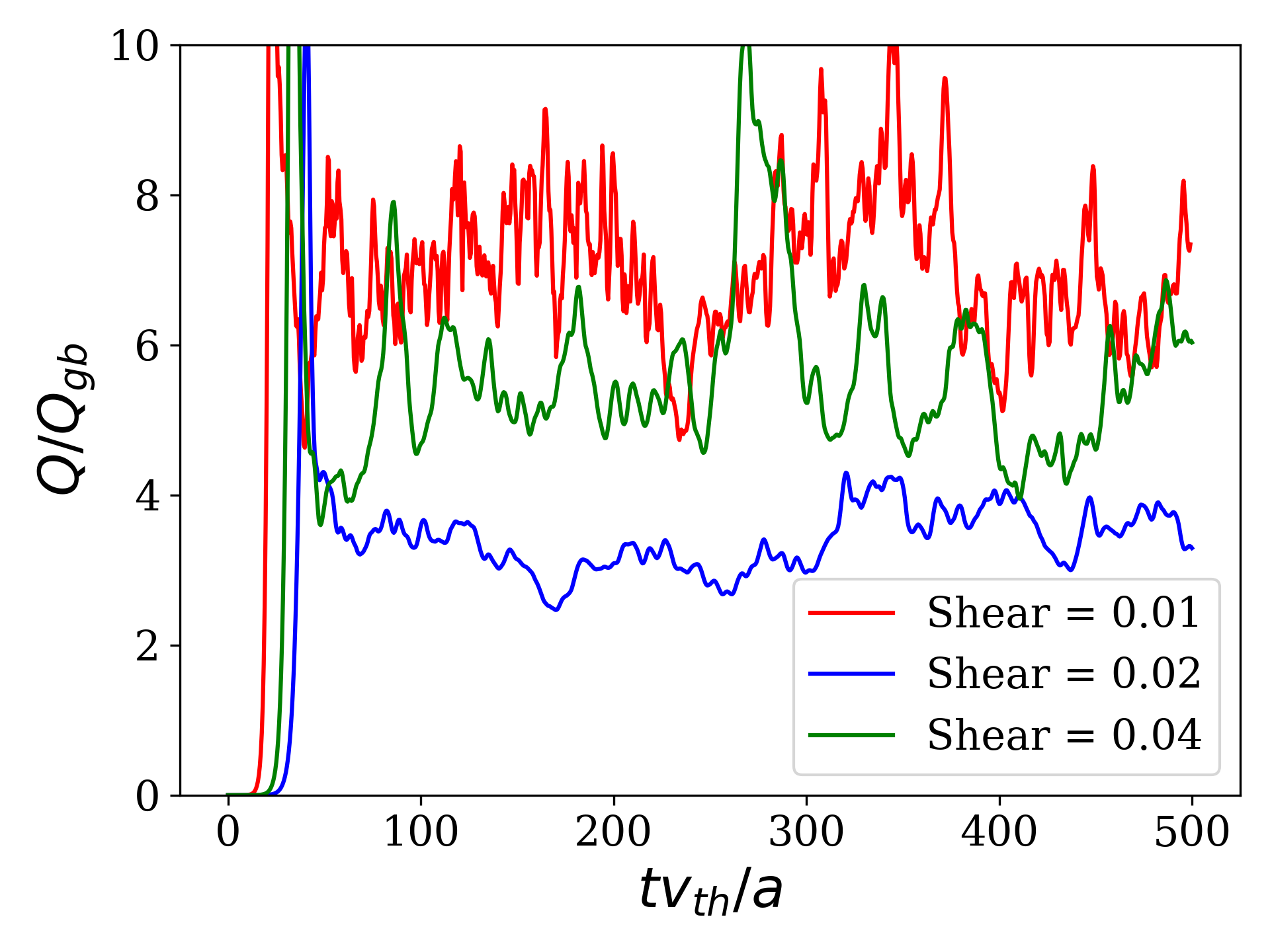}
    \caption{Nonlinear heat flux traces for quasisymmetric equilibria with different shear targets.}
    \label{fig:shear scan}
\end{figure}

To further investigate these effects, we optimized for several more precisely quasisymmetric equilibria, but with different target shears. The nonlinear heat flux traces are shown in Fig. \ref{fig:shear scan}. The shear = 0.01 case is when there is no shear target. While increasing the shear slightly does reduce the heat flux, continuing to increase it eventually increases the heat flux again, consistent with the findings in \citet{Kotschenreuther1995QuantitativeEffects}. It is somewhat surprising that the heat flux is this sensitive to the shear since the shear is very small. For comparison, the global magnetic shear for W7-X in these units is approximately 0.16. Future work will better address the relationship between shear and turbulence in stellarators. This relationship can be very important as very low global magnetic shear is characteristic of vacuum quasisymmetric stellarators \citep{Landreman2022MagneticConfinement}. While the shear increases for finite $\beta$ QS stellarators \citep{Landreman2022OptimizationConfinement}, it is still small compared to shear in tokamaks.

\section{Transport Simulations}\label{trinity sim}
To study how the changes in nonlinear heat fluxes affected the macroscopic profiles, we use the {\tt T3D} transport code \citep{Qian2022StellaratorGX}. {\tt T3D} uses the same algorithm as {\tt Trinity} \citep{Barnes2010DirectCodes}, but is written in Python and adapted to work with stellarator geometries. {\tt T3D} takes advantage of the separation of scales in the local $\delta f$ gyrokinetic model where the distribution function is written as $F = F_{0s} + \delta f_s$. It can be shown that $F_{0s}$ is a Maxwellian and it is assumed that the perturbation $\delta f$ scales like $\delta f \sim \epsilon F_{0s}$ where $\epsilon = \rho/a$. The Maxwellian $F_{0s}$ evolves slower than the perturbation $\partial F_{0s}/\partial t \sim \epsilon^2 \partial \delta f_s/\partial t$. This allows for {\tt T3D} to evolve the macroscopic density, pressure, and temperature profiles using heat and particle fluxes computed by gyrokinetic codes like {\tt GX}.

\begin{figure}
     \centering
     \begin{subfigure}[c]{0.45\textwidth}
         \centering
         \includegraphics[width=\textwidth]{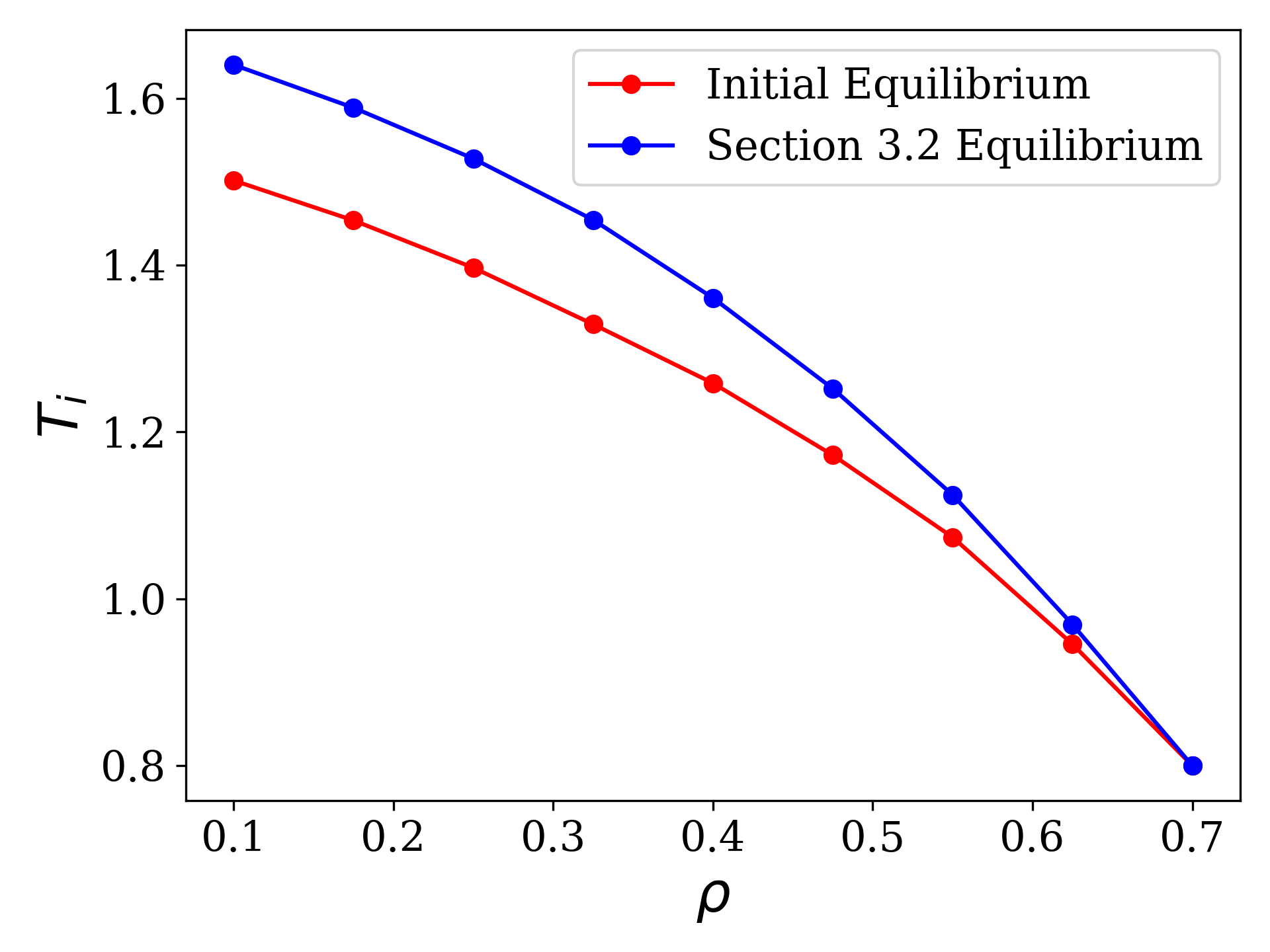}
         \label{temp_comp}
     \end{subfigure}
     \hfill
     \begin{subfigure}[c]{0.45\textwidth}
         \centering
         \includegraphics[width=\textwidth]{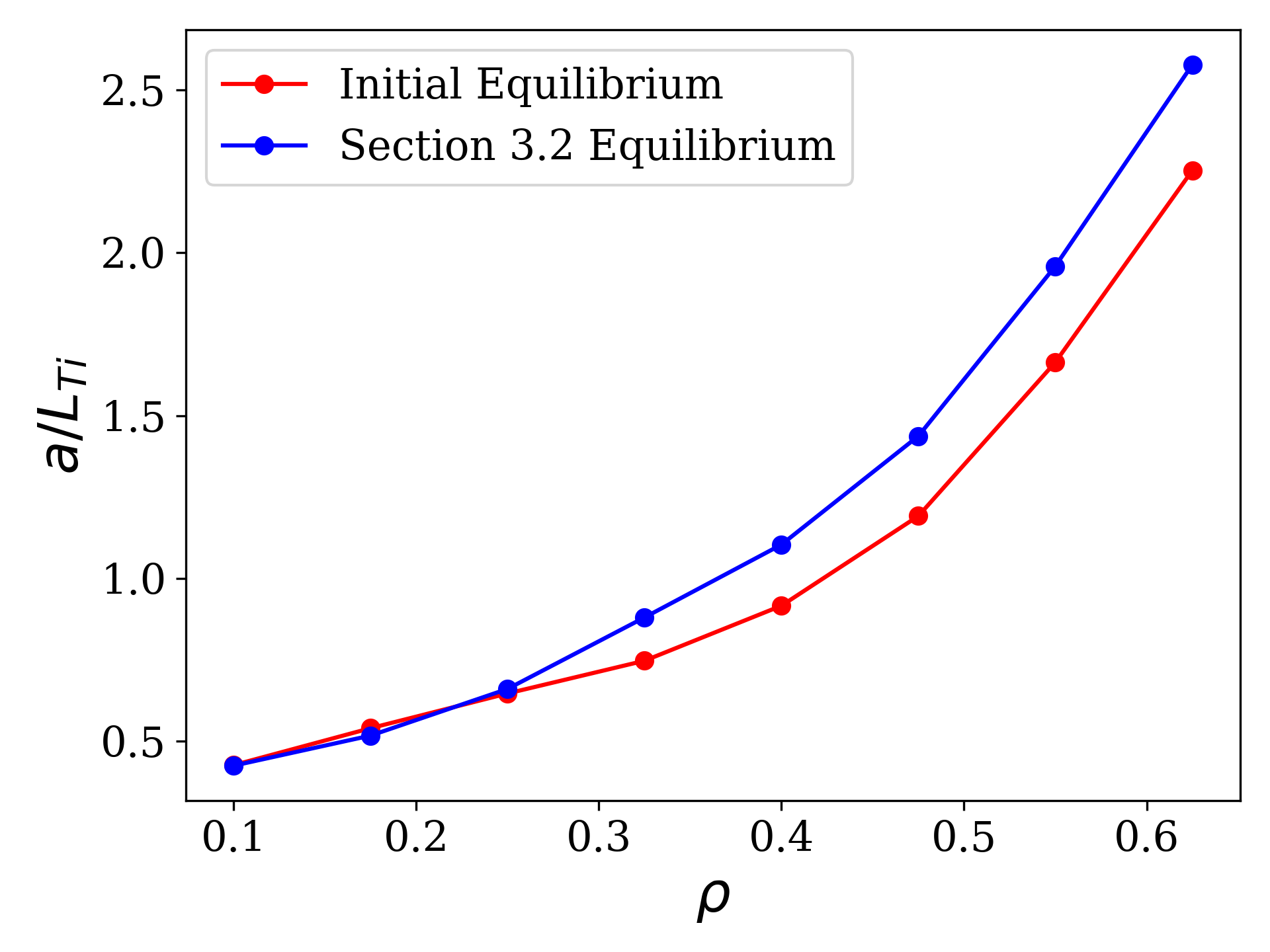}
         \label{temp_grad_comp}
     \end{subfigure}
     \caption{The final temperature (left) and temperature gradient (right) profiles for the initial and optimized equilibria.}
     \label{trinity}
\end{figure}

We run {\tt T3D} simulations for the initial approximately QH equilibrium and optimized equilibrium from Section \ref{turbulence-qs}. Both equilibria were scaled to the same major radius and on-axis magnetic field as W7-X. An adiabatic response is still assumed for the electrons, and neoclassical contributions are ignored to isolate the effects from turbulence. The plots of the final steady-state temperature profiles for both equilibria are shown in Figure \ref{trinity}.

From the plot of the temperature profiles, there is a clear increase in the temperature for the optimized equilibrium. The temperature at the innermost simulated point increased by approximately 10$\%$. This seems lower than expected considering that the optimized equilibrium had lower heat fluxes by about a factor of 3. The true heat flux scales like $Q \sim Q_{sim} n T v_{th} (\rho/a)^2$, where $Q_{sim}$ is the heat flux from the simulation. Since $v_{th} \sim T^{1/2}$ and $\rho \sim v_{th}$, the heat flux scales like $Q \sim Q_{sim} T^{5/2}$. For equivalent sources, $T \sim Q_{sim}^{2/5}$. Therefore, we would expect the temperature to increase by a factor of $Q^{0.4} \sim 1.55$. However, the simulations from Section \ref{turbulence-qs} assumed a temperature gradient of $a/L_{Ti} = 3$. From the plot of the temperature gradients, most points are simulated with lower gradients, and the equilibria have the same critical temperature gradient (as seen in Figure \ref{a/Lt scan}). Nevertheless, the improvement is still very encouraging. Future work will investigate using simulations to optimize for a higher critical gradient or for specific target profiles.

\section{Conclusions}\label{Conclusions}
In this work, we directly optimized stellarators for reduced nonlinear heat fluxes and good quasisymmetry by coupling {\tt GX} and {\tt DESC}. By directly running nonlinear simulations we include the nonlinear saturation mechanisms that determine the steady-state heat flux, and so avoid potential limitations of linear or quasilinear models. The SPSA method is used to handle the noisy heat flux objective and to cheaply estimate the gradient. The newly optimized equilibria show factors of 2-4 improvement in the nonlinear heat flux across several flux surfaces and multiple field lines while also having comparable or improved quasisymmetry. We also investigated how global magnetic shear affects the nonlinear heat flux and found that slightly increasing it can significantly reduce the flux. However, increasing the shear too much led to increases in the heat flux. Our {\tt T3D} simulations showed that by reducing the heat fluxes, the steady-state temperature profiles did increase slightly. Unfortunately, the transport is sufficiently stiff so that macroscopic profiles will approach the critical gradient. Since the initial and optimized equilibria have similar critical gradients, this limits the improvements in the temperature profile. Nevertheless, these results demonstrate that we can efficiently include nonlinear gyrokinetic simulations within the optimization loop. Future work will include adding an objective for MHD stability, optimizing for the nonlinear critical gradient and also directly optimize for desired macroscopic profiles. More improvements will also be made to the SPSA optimizer to make it more effective near minima. Currently, near the minima the true gradient is small enough to be comparable to the simulation noise. This leads to poorer convergence near the minima, and so requires more sophisticated optimization methods. Consequently, we will implement other optimization methods that have been effective for stochastic optimization, such as Bayesian optimization.

\section{Acknowledgements}
The authors thank M. Zarnstoff and B. Buck for insightful and fruitful conversations. Research support came from the U.S Department of Energy (DOE). P.K and colleagues at UMD were supported by the DOE via the Scientific Discovery Through Advanced Computing Program under award number DE-SC0018429 as well as under award number DE-FG0293ER54197. P.K has also been supported through the DOE CSGF Program under award number DE-SC0024386 (while at Princeton University). R.C., D.W.D., E.K., and D.P. were supported by the U.S. Department of Energy under contract numbers DE-AC02-09CH11466,
DE-SC0022005 and Field Work Proposal No. 1019. This work started as part of P.K.'s Summer 2022 DOE SULI internship under award number DE-AC02-09CH11466. 

N.R.M was supported by the DOE Fusion Energy Sciences Postdoctoral Research Program administered by the Oak Ridge Institute for Science and Education (ORISE) for the DOE via Oak Ridge
Associated Universities (ORAU) under DOE contract number DE-SC0014664, and by the Laboratory Directed Research and Development Program of the Princeton Plasma Physics Laboratory under U.S. Department of Energy contract number DE-AC02-09CH11466.

R J is supported by the Portuguese FCT—Fundação para a Ciência e Tecnologia, under Grant 2021.02213.CEECIND. This work has been carried out within the framework of the EUROfusion Consortium, funded by the European Union via the Euratom Research and Training Programme (Grant Agreement No 101052200 — EUROfusion). Views and opinions expressed are however those of the author(s) only and do not necessarily reflect those of the European Union or the European Commission. Neither the European Union nor the European Commission can be held responsible for them. Instituto Superior Técnico activities also received financial support from FCT through Projects UIDB/50010/2020 and UIDP/50010/2020.

Computations were performed on the Traverse and Stellar clusters at Princeton/PPPL as well as the Perlmutter cluster at NERSC.
\appendix
\section{Codes}\label{Codes}
\subsection{{\tt GX}}
{\tt GX} employs the radially-local $\delta\!f$ approach to solve the gyrokinetic equation. In this approximation, the distribution function $F_s$ is represented as $F_s = F_{0s} + \delta\! f_s = F_{Ms}\left(1 - Z_s \Phi/\tau_s\right) + h_s$, where $F_{0s}$ is Maxwellian and $\delta \! f_s$ is a perturbation consisting of a Boltzmann part proprtional to the electrostatic potential $\Phi$ and a general perturbation $h_s$. The subscript $s$ labels species. Assuming electrostatic fluctuations, the gyroaveraged perturbation $g_s = \langle f_s \rangle$ then obeys the electrostatic gyrokinetic equation

\begin{align}\label{gyrokinetic eq}
   \frac{\partial g_s}{\partial t} + & v_{\parallel} \mathbf{b} \cdot \nabla z \left(\frac{\partial g_s}{\partial t} + \frac{q_s}{T_s}\frac{\partial \langle \phi \rangle}{\partial t} F_s\right) - \frac{\mu}{m_s} \mathbf{b}\cdot \nabla z \frac{\partial B}{\partial z}\frac{\partial g_s}{\partial v_{\parallel}} \\
   &+ \mathbf{v}_{Ms} \cdot \left(\nabla_{\perp} g_s + \frac{q_s}{T_s} \nabla_{\perp} \langle \phi \rangle F_s\right)\\
   &+ \langle \mathbf{v}_E \rangle \cdot \nabla_{\perp} g_s + \langle \mathbf{v}_E \rangle \cdot \nabla F_s = \langle C\left(\delta f_s\right)\rangle,
\end{align}
where $v_{\parallel}$ is the velocity parallel to the magnetic field, $\mathbf{b} = \mathbf{B}/B$, $\mu$ is the magnetic moment, $z$ is some coordinate along the field line, $\kappa$ is the field line curvature, $\mathbf{v}_{Ms}$ is the sum of the magnetic and curvature drift velocities, and $\mathbf{v_E}$ is the $\mathbf{E} \times \mathbf{B}$ velocity. For this work, electromagnetic effects are ignored and a Boltzmann response is used to model the perturbed electron distribution.

In {\tt GX}, Eq. \eqref{gyrokinetic eq} is projected onto the Hermite and Laguerre basis functions
\begin{align}
    \psi^l(\mu B) &= (-1)^l e^{-\mu B} \mathrm{L}_l(\mu B), \\
    \phi^m(v_{\parallel}) &= \frac{e^{-v_{\parallel}^2/2}\mathrm{He}_m(v_{\parallel})}{\sqrt{(2 \pi)^3 m!}},
\end{align}
where $\mathrm{He}_m(x)$ and $\mathrm{L}_l(x)$ are the (probabilist's) Hermite and Laguerre polynomials, respectively. More details on the expansion and numerical algorithm can be found in \citet{Mandell2018Laguerre-HermiteGyrokinetics} and \citet{Mandell2022GX:Design}.

By choosing a Hermite-Laguerre basis, the resulting equations for the spectral coefficients reduce to the gyrofluid equations at low resolution \citep{Dorland1993GyrofluidEffects,Beer1995GyrofluidTokamaks}. Particle number, momentum, and energy are also conserved at low resolution. Overall, any inaccuracies are due to the closure or dissipation model used. Therefore, {\tt GX} allows for lower velocity resolution than other similar codes like {\tt GS2} \citep{Kotschenreuther1995ComparisonInstabilities,Dorland2000ElectronTurbulence} and {\tt stella} \citep{Barnes2019Stella:Configurations} that use finite-difference methods in velocity space.  Flexible velocity resolution combined with a GPU implementation allows {\tt GX} to run nonlinear (electrostatic with adiabatic electrons) gyrokinetic simulations in minutes rather than hours or days.
\subsection{{\tt DESC}}
{\tt DESC} \citep{Dudt2020DESC:Solver,Panici2023TheComputations,Conlin2023TheMethods,Dudt2023TheOptimization} is a new stellarator equilibrium and optimization code. Two of the main features of {\tt DESC} are its pseudo-spectral representation of the magnetic geometry and the use of automatic differentiation to compute exact derivatives.

{\tt DESC} directly solves the ideal MHD force balance equation
\begin{equation}
    \mathbf{J} \times \mathbf{B} = \nabla p.
\end{equation}
{\tt DESC} uses as its computational domain the coordinates $(\rho,\theta,\zeta)$, defined as
\begin{subequations}
    \begin{align}
    \rho &= \sqrt{\frac{\psi}{\psi_a}} \\
    \theta &= \theta^* - \lambda(\rho,\theta,\zeta) \\
    \zeta &= \phi,
\end{align}
\end{subequations}
where $\psi$ is the enclosed toroidal flux, $\psi_a$ is the total enclosed toroidal flux in the plasma, $\theta^*$ is a straight field line poloidal angle, $\lambda$ is a stream function, and $\phi$ is the geometric toroidal angle. It then expands $\lambda$, and the cylindrical coordinates $R$ and $Z$ in terms of a Fourier-Zernike basis.

\begin{subequations}
\begin{equation}
    R(\rho,\theta,\zeta) = \sum_{m=-M,n=-N,l=0}^{M,N,L} R_{lmn} \mathcal{Z}_l^m (\rho,\theta) \mathcal{F}^n(\zeta)
\end{equation}
\begin{equation}
    \lambda(\rho,\theta,\zeta) = \sum_{m=-M,n=-N,l=0}^{M,N,L} \lambda_{lmn} \mathcal{Z}_l^m (\rho,\theta) \mathcal{F}^n(\zeta)
\end{equation}
\begin{equation}
Z(\rho,\theta,\zeta) = \sum_{m=-M,n=-N,l=0}^{M,N,L} Z_{lmn} \mathcal{Z}_l^m (\rho,\theta) \mathcal{F}^n(\zeta)
\end{equation}
\end{subequations}
where $\mathcal{R}_l^{|m|}$, $\mathcal{Z}_l^m$, and $\mathcal{F}$ are the shifted Jacobi polynomials, Zernike polynomials, and Fourier series, respectively.

It can be shown that the force error $\mathbf{J} \times \mathbf{B} - \nabla p$ has only two independent components
\begin{align}
    F_\rho &= \sqrt{g}\left(J^{\zeta}B^{\theta} - J^{\theta}B^{\zeta}\right) + p' \\
    F_\beta &= \sqrt{g}J^{\rho},
\end{align}
where $\sqrt{g}$ is the Jacobian, and the superscripts indicate the contravariant components. {\tt DESC} then solves for the spectral coefficients $R_{lmn}$, $Z_{lmn}$, and $\lambda_{lmn}$ that minimize this force error.

{\tt DESC} also utilizes automatic differentiation through JAX \citep{jax2018github} in its optimization routines. Briefly, in automatic differentiation, the chain rule is applied through the code, allowing {\tt DESC} to compute exact derivatives faster and more accurately than derivatives from finite differencing routines. However, at the time of this writing this requires writing the objective function in native Python, and {\tt GX} is written in C++/CUDA. Furthermore, it is unclear how effective it would be to differentiate through the timestepping of such a noisy system. Therefore, this would require extensive code developments for both {\tt DESC} and {\tt GX}, and so AD will not be used when optimizing for reduced turbulence.
\section{Simulation Parameters used in Optimization}\label{app sim param}
\subsection{Turbulence, Turbulence-QS, and Multiple $\alpha$ Optimization}
For the {\tt GX} simulations within the optimization loop, we use the simulation parameters listed in Table \ref{sim table}.
\begin{table}
\centering
    \begin{tabular}{cc}
     \hline
 Parameter & Value\\ [0.5ex] 
 \hline\hline
 Normalized Toroidal Flux (s) & 0.5\\
 \hline
 field line label ($\alpha$) & 0.0 (and $\pi \iota/4)$ for multiple $\alpha$\\
 \hline
Number of Poloidal Turns (npol) & 1\\
\hline
Parallel Resolution (ntheta) & 64 \\
\hline
Radial resolution (nx)& 64 \\
\hline
field line label resolution (ny) & 64 \\
\hline
Hermite Resolution (nhermite) & 8 \\
\hline
Laguerre Resolution (nlaguerre) & 4\\
\hline
Normalized Temperature Gradient (tprim) & 3.0 \\
\hline
Normalized Density Gradient (fprim) & 1.0 \\
\hline
    \end{tabular}
\caption{The simulation parameters used for the {\tt GX} simulations within the optimization loop for Sections \ref{turbulence only opt}-\ref{multiple alpha case}.}
\label{sim table}
\end{table}

The number of poloidal turns, parallel resolution, and number of simulated modes (proportional to $n_x$ and $n_y$) are chosen to enable cheaper simulations. However, one poloidal turn has been used in previous nonlinear W7-X gyrokinetic benchmark studies \citep{Gonzalez-Jerez2022ElectrostaticGENEb}. Furthermore, from a quasilinear estimate the heat flux scales like $1/\langle k_{\perp}^2\rangle$ (where the angle brackets indicate a flux-surface average) \citep{Mariani2018IdentifyingFluxes}. Consequently, higher wave number modes contribute less to the heat flux. The gradients are typical for experimental profiles for W7-X \citep{Beurskens2021IonPlasmas} and have been previously used for benchmark, transport, and optimization studies \citep{Gonzalez-Jerez2022ElectrostaticGENE, BanonNavarro2023First-principlesStellarators, Roberg-Clark2022ReductionOptimization}. Finally, it's been demonstrated that {\tt GX} simulations can yield accurate heat flux traces using the specified number of Hermite and Laguerre polynomials \citep{Mandell2022GX:Design}. To check our results, the post-processing simulations are run at higher resolution.

\subsection{Fluid Approximation Optimization}
When running the optimization at fluid resolution, the simulation parameters are identical except those listed in Table \ref{sim table fluid}.
\begin{table}
\centering
    \begin{tabular}{cc}
     \hline
 Parameter & Value\\ [0.5ex] 
 \hline\hline
Hermite Resolution (nhermite) & 4 \\
\hline
Laguerre Resolution (nlaguerre) & 2\\
\hline
Collision Frequency (vnewk) & 2.0\\
\hline
Normalized Temperature Gradient (tprim) & 5.0 \\
\hline
Normalized Density Gradient (fprim) & 1.0 \\
\hline
    \end{tabular}
\caption{The simulation parameters used for the {\tt GX} simulations within the optimization loop for the fluid case in Section \ref{fluid opt}}
\label{sim table fluid}
\end{table}
The velocity resolution is decreased to that of the gyrofluid model, and the collision frequency is increased. Since there is far greater dissipation in the model, we also increase the temperature gradient.

\subsection{Post-processing Simulations}
To ensure our final results are well-converged, we increase the resolution of the post-processing simulations to those listed in Table \ref{sim table post}.
\begin{table}
\centering
    \begin{tabular}{cc}
     \hline
 Parameter & Value\\ [0.5ex] 
 \hline\hline
Number of Poloidal Turns (npol) & 2\\
\hline
Parallel Resolution (ntheta) & 128 \\
\hline
Radial resolution (nx)& 128 \\
\hline
field line label resolution (ny) & 128 \\
\hline
Hermite Resolution (nhermite) & 16 \\
\hline
Laguerre Resolution (nlaguerre) & 8\\
\hline
    \end{tabular}
\caption{The simulation parameters used for the {\tt GX} simulations for post-processing.}
\label{sim table post}
\end{table}
The scans over $\rho$, $\alpha$, and $a/L_T$ of course include additional values.

\subsection{{\tt DESC} Equilibria Parameters}
The parameters used for the equilibria are shown in this table. These are considered "moderate" resolution, and are typical values used for optimization. This is a vacuum case, so both the pressure and current are zero. When optimized for precise QH symmetry, the resulting equilibrium is similar to the QH equilibirum in \citep{Landreman2022MagneticConfinement}.
\begin{table}
\centering
    \begin{tabular}{cc}
     \hline
 Parameter & Value\\ [0.5ex] 
 \hline\hline
Spectral Resolution (M, N, L) & (8, 8, 8)\\
\hline
Grid Resolution (M$_{grid}$, L$_{grid}$, N$_{grid}$) & (16, 16, 16) \\
\hline
Total Toroidal Flux (Psi) & 0.03817902 \\
\hline
Major Radius (R0) & 1.0 \\
\hline
Aspect Ratio (R0/a) & 8.0 \\
\hline
    \end{tabular}
\caption{The resolution parameters for the {\tt DESC} equilibria in this study.}
\label{desc parameters}
\end{table}


\subsection{Optimization Timings}
The table below shows the approximate timings of different parts of the optimization loop using a single Tesla V100 GPU on the Princeton Traverse cluster. All values are in minutes. In total, each optimization completed in about 20 hours. It takes slightly more than 12 hours on a single NVIDIA A100 GPU on the NERSC Perlmutter cluster. The "optimization" timings are for a single iteration.
\begin{table}
\centering
    \begin{tabular}{cc}
     \hline
 Parameter & Value\\ [0.5ex] 
 \hline\hline
{\tt GX} Evaluation & 3\\
\hline
{\tt GX} Optimization & 10 \\
\hline
Quasisymmetry Optimization & 4 \\
\hline
    \end{tabular}
\caption{The approximate timings for different parts of the optimization loop.}
\label{timings}
\end{table}
\bibliographystyle{jpp}

\newpage

\section{DESC Input Files}
\subsection{Initial Equilibrium}
\begin{verbatim}
# DESC input file generated from the output file:
# test_equilibria/traverse/input.nfp4_QH_desc_output.h5

# global parameters
sym = 1
NFP = 4
Psi = 3.81790200E-02

# spectral resolution
L_rad = 8
M_pol = 8
N_tor = 8
L_grid = 16
M_grid = 16
N_grid = 16

# solver tolerances
ftol = 1.000E-02
xtol = 1.000E-06
gtol = 1.000E-06
maxiter = 100

# solver methods
optimizer = lsq-exact
objective = force
spectral_indexing = ansi

# pressure and rotational transform/current profiles
l:   0  p =  0.00000000E+00  c =  6.27347854E-14
l:   1  p =  0.00000000E+00  c =  1.71917298E-10
l:   2  p =  0.00000000E+00  c = -5.44257894E-09
l:   3  p =  0.00000000E+00  c =  5.92952658E-08
l:   4  p =  0.00000000E+00  c = -2.95964833E-07
l:   5  p =  0.00000000E+00  c =  7.63421786E-07
l:   6  p =  0.00000000E+00  c = -1.05372308E-06
l:   7  p =  0.00000000E+00  c =  7.38152073E-07

# fixed-boundary surface shape
l:   0  m:  -3  n:  -3  R1 =  1.06009580E-04  Z1 =  0.00000000E+00
l:   0  m:  -2  n:  -3  R1 =  1.94431471E-03  Z1 =  0.00000000E+00
l:   0  m:  -1  n:  -3  R1 = -4.45641661E-03  Z1 =  0.00000000E+00
l:   0  m:  -3  n:  -2  R1 = -1.41499877E-03  Z1 =  0.00000000E+00
l:   0  m:  -2  n:  -2  R1 =  3.96888996E-03  Z1 =  0.00000000E+00
l:   0  m:  -1  n:  -2  R1 = -3.13235773E-02  Z1 =  0.00000000E+00
l:   0  m:  -3  n:  -1  R1 = -4.34044222E-04  Z1 =  0.00000000E+00
l:   0  m:  -2  n:  -1  R1 =  2.23464111E-02  Z1 =  0.00000000E+00
l:   0  m:  -1  n:  -1  R1 = -1.16349515E-01  Z1 =  0.00000000E+00
l:   0  m:   0  n:   0  R1 =  1.00000000E+00  Z1 =  0.00000000E+00
l:   0  m:   1  n:   0  R1 =  1.38039047E-01  Z1 =  0.00000000E+00
l:   0  m:   2  n:   0  R1 =  1.53605591E-02  Z1 =  0.00000000E+00
l:   0  m:   3  n:   0  R1 =  1.06941717E-04  Z1 =  0.00000000E+00
l:   0  m:   0  n:   1  R1 =  1.79819788E-01  Z1 =  0.00000000E+00
l:   0  m:   1  n:   1  R1 = -5.82857335E-02  Z1 =  0.00000000E+00
l:   0  m:   2  n:   1  R1 =  1.99922621E-02  Z1 =  0.00000000E+00
l:   0  m:   3  n:   1  R1 =  5.62751826E-04  Z1 =  0.00000000E+00
l:   0  m:   0  n:   2  R1 =  1.73495115E-02  Z1 =  0.00000000E+00
l:   0  m:   1  n:   2  R1 = -1.75730831E-02  Z1 =  0.00000000E+00
l:   0  m:   2  n:   2  R1 =  6.57072137E-03  Z1 =  0.00000000E+00
l:   0  m:   3  n:   2  R1 = -7.49945379E-04  Z1 =  0.00000000E+00
l:   0  m:   0  n:   3  R1 =  2.57716016E-03  Z1 =  0.00000000E+00
l:   0  m:   1  n:   3  R1 = -2.82630336E-03  Z1 =  0.00000000E+00
l:   0  m:   2  n:   3  R1 =  1.99926226E-03  Z1 =  0.00000000E+00
l:   0  m:   3  n:   3  R1 = -6.16014311E-05  Z1 =  0.00000000E+00
l:   0  m:   0  n:  -3  R1 =  0.00000000E+00  Z1 = -1.19971129E-03
l:   0  m:   1  n:  -3  R1 =  0.00000000E+00  Z1 =  2.92305608E-03
l:   0  m:   2  n:  -3  R1 =  0.00000000E+00  Z1 = -3.11190235E-03
l:   0  m:   3  n:  -3  R1 =  0.00000000E+00  Z1 =  6.31026425E-04
l:   0  m:   0  n:  -2  R1 =  0.00000000E+00  Z1 = -1.86633423E-02
l:   0  m:   1  n:  -2  R1 =  0.00000000E+00  Z1 =  1.74756774E-02
l:   0  m:   2  n:  -2  R1 =  0.00000000E+00  Z1 = -8.13554668E-03
l:   0  m:   3  n:  -2  R1 =  0.00000000E+00  Z1 =  4.05079520E-04
l:   0  m:   0  n:  -1  R1 =  0.00000000E+00  Z1 = -1.52687429E-01
l:   0  m:   1  n:  -1  R1 =  0.00000000E+00  Z1 =  1.24701868E-02
l:   0  m:   2  n:  -1  R1 =  0.00000000E+00  Z1 =  1.90954690E-02
l:   0  m:   3  n:  -1  R1 =  0.00000000E+00  Z1 =  4.56955309E-04
l:   0  m:  -3  n:   0  R1 =  0.00000000E+00  Z1 = -3.80285523E-04
l:   0  m:  -2  n:   0  R1 =  0.00000000E+00  Z1 = -9.45114014E-03
l:   0  m:  -1  n:   0  R1 =  0.00000000E+00  Z1 = -1.31096801E-01
l:   0  m:  -3  n:   1  R1 =  0.00000000E+00  Z1 = -1.82574602E-03
l:   0  m:  -2  n:   1  R1 =  0.00000000E+00  Z1 = -1.82165803E-02
l:   0  m:  -1  n:   1  R1 =  0.00000000E+00  Z1 = -7.78740189E-02
l:   0  m:  -3  n:   2  R1 =  0.00000000E+00  Z1 = -8.84087119E-04
l:   0  m:  -2  n:   2  R1 =  0.00000000E+00  Z1 =  6.09077798E-03
l:   0  m:  -1  n:   2  R1 =  0.00000000E+00  Z1 = -3.10828300E-02
l:   0  m:  -3  n:   3  R1 =  0.00000000E+00  Z1 = -6.25309212E-04
l:   0  m:  -2  n:   3  R1 =  0.00000000E+00  Z1 =  3.23790883E-03
l:   0  m:  -1  n:   3  R1 =  0.00000000E+00  Z1 = -4.64192523E-03
\end{verbatim}

\subsection{Optimized Equilibrium from Section \ref{turbulence-qs}}
\begin{verbatim}
# DESC input file generated from the output file:
# test_equilibria/traverse/swap_gx_end_case23_n_3_i_14.h5

# global parameters
sym = 1
NFP = 4
Psi = 3.81790200E-02

# spectral resolution
L_rad = 8
M_pol = 8
N_tor = 8
L_grid = 16
M_grid = 16
N_grid = 16

# solver tolerances
ftol = 1.000E-02
xtol = 1.000E-06
gtol = 1.000E-06
maxiter = 100

# solver methods
optimizer = lsq-exact
objective = force
spectral_indexing = ansi

# pressure and rotational transform/current profiles
l:   0  p =  0.00000000E+00  c =  1.41645009E-12
l:   1  p =  0.00000000E+00  c = -9.25103059E-10
l:   2  p =  0.00000000E+00  c =  2.19561377E-08
l:   3  p =  0.00000000E+00  c = -1.86967349E-07
l:   4  p =  0.00000000E+00  c =  7.72543397E-07
l:   5  p =  0.00000000E+00  c = -1.72630498E-06
l:   6  p =  0.00000000E+00  c =  2.13214903E-06
l:   7  p =  0.00000000E+00  c = -1.36872255E-06

# fixed-boundary surface shape
l:   0  m:  -3  n:  -3  R1 = -9.31357764E-03  Z1 =  0.00000000E+00
l:   0  m:  -2  n:  -3  R1 =  5.71628208E-03  Z1 =  0.00000000E+00
l:   0  m:  -1  n:  -3  R1 =  3.82681062E-03  Z1 =  0.00000000E+00
l:   0  m:  -3  n:  -2  R1 = -2.29388344E-03  Z1 =  0.00000000E+00
l:   0  m:  -2  n:  -2  R1 =  1.66286469E-02  Z1 =  0.00000000E+00
l:   0  m:  -1  n:  -2  R1 = -2.90057270E-02  Z1 =  0.00000000E+00
l:   0  m:  -3  n:  -1  R1 =  5.54126280E-03  Z1 =  0.00000000E+00
l:   0  m:  -2  n:  -1  R1 = -5.52415883E-03  Z1 =  0.00000000E+00
l:   0  m:  -1  n:  -1  R1 = -9.11377065E-02  Z1 =  0.00000000E+00
l:   0  m:   0  n:   0  R1 =  1.00000000E+00  Z1 =  0.00000000E+00
l:   0  m:   1  n:   0  R1 =  1.22407301E-01  Z1 =  0.00000000E+00
l:   0  m:   2  n:   0  R1 =  1.27821030E-02  Z1 =  0.00000000E+00
l:   0  m:   3  n:   0  R1 = -4.34547934E-03  Z1 =  0.00000000E+00
l:   0  m:   0  n:   1  R1 =  1.87116351E-01  Z1 =  0.00000000E+00
l:   0  m:   1  n:   1  R1 = -3.72058062E-02  Z1 =  0.00000000E+00
l:   0  m:   2  n:   1  R1 = -1.28924376E-02  Z1 =  0.00000000E+00
l:   0  m:   3  n:   1  R1 = -3.16332008E-03  Z1 =  0.00000000E+00
l:   0  m:   0  n:   2  R1 =  1.98088376E-02  Z1 =  0.00000000E+00
l:   0  m:   1  n:   2  R1 = -1.77228005E-02  Z1 =  0.00000000E+00
l:   0  m:   2  n:   2  R1 =  1.84202872E-02  Z1 =  0.00000000E+00
l:   0  m:   3  n:   2  R1 = -4.06745395E-03  Z1 =  0.00000000E+00
l:   0  m:   0  n:   3  R1 = -2.60990147E-04  Z1 =  0.00000000E+00
l:   0  m:   1  n:   3  R1 =  6.53196582E-04  Z1 =  0.00000000E+00
l:   0  m:   2  n:   3  R1 =  5.67484731E-03  Z1 =  0.00000000E+00
l:   0  m:   3  n:   3  R1 = -8.14020474E-03  Z1 =  0.00000000E+00
l:   0  m:   0  n:  -3  R1 =  0.00000000E+00  Z1 = -2.70012676E-03
l:   0  m:   1  n:  -3  R1 =  0.00000000E+00  Z1 =  2.98740559E-03
l:   0  m:   2  n:  -3  R1 =  0.00000000E+00  Z1 = -6.43349941E-03
l:   0  m:   3  n:  -3  R1 =  0.00000000E+00  Z1 =  4.61608984E-03
l:   0  m:   0  n:  -2  R1 =  0.00000000E+00  Z1 = -1.16055055E-02
l:   0  m:   1  n:  -2  R1 =  0.00000000E+00  Z1 =  4.49890508E-03
l:   0  m:   2  n:  -2  R1 =  0.00000000E+00  Z1 = -1.21016790E-02
l:   0  m:   3  n:  -2  R1 =  0.00000000E+00  Z1 =  1.27557793E-03
l:   0  m:   0  n:  -1  R1 =  0.00000000E+00  Z1 = -1.07649166E-01
l:   0  m:   1  n:  -1  R1 =  0.00000000E+00  Z1 =  2.76464573E-02
l:   0  m:   2  n:  -1  R1 =  0.00000000E+00  Z1 = -9.28264520E-03
l:   0  m:   3  n:  -1  R1 =  0.00000000E+00  Z1 =  1.51173953E-03
l:   0  m:  -3  n:   0  R1 =  0.00000000E+00  Z1 =  7.72083960E-03
l:   0  m:  -2  n:   0  R1 =  0.00000000E+00  Z1 = -2.60627865E-03
l:   0  m:  -1  n:   0  R1 =  0.00000000E+00  Z1 = -1.56329762E-01
l:   0  m:  -3  n:   1  R1 =  0.00000000E+00  Z1 =  6.64991771E-03
l:   0  m:  -2  n:   1  R1 =  0.00000000E+00  Z1 =  1.41663507E-02
l:   0  m:  -1  n:   1  R1 =  0.00000000E+00  Z1 = -7.37699026E-02
l:   0  m:  -3  n:   2  R1 =  0.00000000E+00  Z1 = -1.91298905E-03
l:   0  m:  -2  n:   2  R1 =  0.00000000E+00  Z1 =  1.48107607E-02
l:   0  m:  -1  n:   2  R1 =  0.00000000E+00  Z1 = -1.58727989E-02
l:   0  m:  -3  n:   3  R1 =  0.00000000E+00  Z1 = -5.73744002E-03
l:   0  m:  -2  n:   3  R1 =  0.00000000E+00  Z1 =  5.43576784E-03
l:   0  m:  -1  n:   3  R1 =  0.00000000E+00  Z1 =  8.48615937E-04
\end{verbatim}
\bibliography{references}

\begin{thebibliography}{49}
\expandafter\ifx\csname natexlab\endcsname\relax\def\natexlab#1{#1}\fi
\def\au#1{#1} \def\ed#1{#1} \def\yr#1{#1}\def\at#1{#1}\def\jt#1{\textit{#1}} \def\bt#1{#1}\def\bvol#1{\textbf{#1}} \def\vol#1{#1} \def\pg#1{#1} \def\publ#1{#1}\def\arxiv#1{#1}\def\org#1{#1}\def\st#1{\textit{#1}}

\bibitem[Antonsen~Jr. \& Lane(1980)]{AntonsenJr.1980KineticPlasmas}
{\sc \au{Antonsen~Jr., Thomas~M} \& \au{Lane, Barton}} \yr{1980}  \at{{Kinetic equations for low frequency instabilities in inhomogeneous plasmas}}.  \jt{The Physics of Fluids}  \bvol{23}~(6),  \pg{1205--1214}.

\bibitem[Bader {\em et~al.\/}(2020)Bader, Faber, Schmitt, Anderson, Drevlak, Duff, Frerichs, Hegna, Kruger, Landreman, McKinney, Singh, Schroeder, Terry \& Ware]{Bader2020AdvancingStellarators}
{\sc \au{Bader, A.}, \au{Faber, B.~J.}, \au{Schmitt, J.~C.}, \au{Anderson, D.~T.}, \au{Drevlak, M.}, \au{Duff, J.~M.}, \au{Frerichs, H.}, \au{Hegna, C.~C.}, \au{Kruger, T.~G.}, \au{Landreman, M.}, \au{McKinney, I.~J.}, \au{Singh, L.}, \au{Schroeder, J.~M.}, \au{Terry, P.~W.} \& \au{Ware, A.~S.}} \yr{2020}  \at{{Advancing the physics basis for quasi-helically symmetric stellarators}}.  \jt{Journal of Plasma Physics}  \bvol{86}~(5),  \pg{905860506}.

\bibitem[Ba{\~{n}}{\'{o}}n~Navarro {\em et~al.\/}(2023)Ba{\~{n}}{\'{o}}n~Navarro, Di~Siena, Velasco, Wilms, Merlo, Windisch, LoDestro, Parker \& Jenko]{BanonNavarro2023First-principlesStellarators}
{\sc \au{Ba{\~{n}}{\'{o}}n~Navarro, A.}, \au{Di~Siena, A.}, \au{Velasco, J.L.}, \au{Wilms, F.}, \au{Merlo, G.}, \au{Windisch, T.}, \au{LoDestro, L.L.}, \au{Parker, J.B.} \& \au{Jenko, F.}} \yr{2023}  \at{{First-principles based plasma profile predictions for optimized stellarators}}.  \jt{Nuclear Fusion}  \bvol{63}~(5),  \pg{054003}.

\bibitem[Barnes {\em et~al.\/}(2010)Barnes, Abel, Dorland, G{\"{o}}rler, Hammett \& Jenko]{Barnes2010DirectCodes}
{\sc \au{Barnes, M.}, \au{Abel, I.~G.}, \au{Dorland, W.}, \au{G{\"{o}}rler, T.}, \au{Hammett, G.~W.} \& \au{Jenko, F.}} \yr{2010}  \at{{Direct multiscale coupling of a transport code to gyrokinetic turbulence codes}}.  \jt{Physics of Plasmas}  \bvol{17}~(5),  \pg{056109}.

\bibitem[Barnes {\em et~al.\/}(2019)Barnes, Parra \& Landreman]{Barnes2019Stella:Configurations}
{\sc \au{Barnes, M.}, \au{Parra, F.~I.} \& \au{Landreman, M.}} \yr{2019}  \at{{stella: An operator-split, implicit–explicit {$\delta$}f-gyrokinetic code for general magnetic field configurations}}.  \jt{Journal of Computational Physics}  \bvol{391},  \pg{365--380}.

\bibitem[Beer(1995)]{Beer1995GyrofluidTokamaks}
{\sc \au{Beer, Michael~Alan}} \yr{1995}  \at{{Gyrofluid Models of Turbulent Transport in Tokamaks}}. PhD thesis.

\bibitem[Beidler {\em et~al.\/}(2021)Beidler, Smith, Alonso, Andreeva, Baldzuhn, A~Beurskens, Borchardt, Bozhenkov, Brunner, Damm, Drevlak, Ford, Fuchert, Geiger, Helander, Hergenhahn, Hirsch, H{\"{o}}fel, Kazakov, Kleiber, Krychowiak, Kwak, Langenberg, Laqua, Neuner, Pablant, Pasch, Pavone, Pedersen, Rahbarnia, Schilling, Scott, Stange, Svensson, Thomsen, Turkin, Warmer, Wolf \& Zhang]{Beidler2021Demonstration7-X}
{\sc \au{Beidler, C~D}, \au{Smith, H~M}, \au{Alonso, A}, \au{Andreeva, T}, \au{Baldzuhn, J}, \au{A~Beurskens, M~N}, \au{Borchardt, M}, \au{Bozhenkov, S~A}, \au{Brunner, K~J}, \au{Damm, H}, \au{Drevlak, M}, \au{Ford, O~P}, \au{Fuchert, G}, \au{Geiger, J}, \au{Helander, P}, \au{Hergenhahn, U}, \au{Hirsch, M}, \au{H{\"{o}}fel, U}, \au{Kazakov, Ye~O}, \au{Kleiber, R}, \au{Krychowiak, M}, \au{Kwak, S}, \au{Langenberg, A}, \au{Laqua, H~P}, \au{Neuner, U}, \au{Pablant, N~A}, \au{Pasch, E}, \au{Pavone, A}, \au{Pedersen, T~S}, \au{Rahbarnia, K}, \au{Schilling, J}, \au{Scott, E~R}, \au{Stange, T}, \au{Svensson, J}, \au{Thomsen, H}, \au{Turkin, Y}, \au{Warmer, F}, \au{Wolf, R~C} \& \au{Zhang, D}} \yr{2021}  \at{{Demonstration of reduced neoclassical energy transport in Wendelstein 7-X}}.  \jt{Nature}  \bvol{596},  \pg{221–226}.

\bibitem[Beurskens {\em et~al.\/}(2021)Beurskens, Bozhenkov, Ford, Xanthopoulos, Zocco, Turkin, Alonso, Beidler, Calvo, Carralero, Estrada, Fuchert, Grulke, Hirsch, Ida, Jakubowski, Killer, Krychowiak, Kwak, Lazerson, Langenberg, Lunsford, Pablant, Pasch, Pavone, Reimold, Romba, von Stechow, Smith, Windisch, Yoshinuma, Zhang, Wolf \& W7-X~Team]{Beurskens2021IonPlasmas}
{\sc \au{Beurskens, M.N.A.}, \au{Bozhenkov, S.A.}, \au{Ford, O.}, \au{Xanthopoulos, P.}, \au{Zocco, A.}, \au{Turkin, Y.}, \au{Alonso, A.}, \au{Beidler, C.}, \au{Calvo, I.}, \au{Carralero, D.}, \au{Estrada, T.}, \au{Fuchert, G.}, \au{Grulke, O.}, \au{Hirsch, M.}, \au{Ida, K.}, \au{Jakubowski, M.}, \au{Killer, C.}, \au{Krychowiak, M.}, \au{Kwak, S.}, \au{Lazerson, S.}, \au{Langenberg, A.}, \au{Lunsford, R.}, \au{Pablant, N.}, \au{Pasch, E.}, \au{Pavone, A.}, \au{Reimold, F.}, \au{Romba, Th.}, \au{von Stechow, A.}, \au{Smith, H.M.}, \au{Windisch, T.}, \au{Yoshinuma, M.}, \au{Zhang, D.}, \au{Wolf, R.C.} \& \au{W7-X~Team, the}} \yr{2021}  \at{{Ion temperature clamping in Wendelstein 7-X electron cyclotron heated plasmas}}.  \jt{Nuclear Fusion}  \bvol{61}~(11),  \pg{116072}.

\bibitem[Bradbury {\em et~al.\/}(2018)Bradbury, Frostig, Hawkins, Johnson, Leary, Maclaurin, Necula, Paszke, VanderPlas, Wanderman-Milne \& Zhang]{jax2018github}
{\sc \au{Bradbury, James}, \au{Frostig, Roy}, \au{Hawkins, Peter}, \au{Johnson, Matthew~James}, \au{Leary, Chris}, \au{Maclaurin, Dougal}, \au{Necula, George}, \au{Paszke, Adam}, \au{VanderPlas, Jake}, \au{Wanderman-Milne, Skye} \& \au{Zhang, Qiao}} \yr{2018} {{\{}JAX{\}}: composable transformations of {\{}P{\}}ython+{\{}N{\}}um{\{}P{\}}y programs}.

\bibitem[Buck {\em et~al.\/}(2022)Buck, Dorland, Mandell, Kim, Fischer \& Qian]{Buck2022AGX}
{\sc \au{Buck, Braden}, \au{Dorland, William}, \au{Mandell, Noah}, \au{Kim, Patrick}, \au{Fischer, Sorah} \& \au{Qian, Tony}} \yr{2022} {A comparison of numerical and analytic ITG turbulence models in the gyrokinetic code GX}.

\bibitem[Buller {\em et~al.\/}(2023)Buller, Mandell, Parisi, Kim, Adkins, Gaur, Dorland \& Landreman]{Buller2023LinearStellarators}
{\sc \au{Buller, Stefan}, \au{Mandell, Noah}, \au{Parisi, Jason}, \au{Kim, Patrick}, \au{Adkins, Toby}, \au{Gaur, Rahul}, \au{Dorland, William} \& \au{Landreman, Matt}} \yr{2023} {Linear Physics Proxies for the Nonlinear Heat Flux in Stellarators}.  \bt{In {\em Simons-CIEMAT Joint Meeting on Stellarator Turbulence Optimization\/}}. Madrid.

\bibitem[Catto(1978)]{Catto1978LinearizedGyro-kinetics}
{\sc \au{Catto, P~J}} \yr{1978}  \at{{Linearized gyro-kinetics}}.  \jt{Plasma Physics}  \bvol{20}~(7),  \pg{719--722}.

\bibitem[Chan {\em et~al.\/}(2003)Chan, Doucet \& Tadic]{Chan2003OptimisationApproximation}
{\sc \au{Chan, Bao~Ling}, \au{Doucet, A} \& \au{Tadic, V~B}} \yr{2003} {Optimisation of particle filters using simultaneous perturbation stochastic approximation}.  \bt{In {\em 2003 IEEE International Conference on Acoustics, Speech, and Signal Processing, 2003. Proceedings. (ICASSP '03).\/}}, ,  \vol{vol.~6},  \pg{pp. VI--681}.

\bibitem[Conlin {\em et~al.\/}(2023)Conlin, Dudt, Panici \& Kolemen]{Conlin2023TheMethods}
{\sc \au{Conlin, Rory}, \au{Dudt, Daniel~W.}, \au{Panici, Dario} \& \au{Kolemen, Egemen}} \yr{2023}  \at{{The DESC stellarator code suite. Part 2. Perturbation and continuation methods}}.  \jt{Journal of Plasma Physics}  \bvol{89}~(3),  \pg{955890305}.

\bibitem[Dewar \& Glasser(1983)]{Dewar1983BallooningSystems}
{\sc \au{Dewar, R~L} \& \au{Glasser, A~H}} \yr{1983}  \at{{Ballooning mode spectrum in general toroidal systems}}.  \jt{The Physics of Fluids}  \bvol{26}~(10),  \pg{3038--3052}.

\bibitem[Dorland \& Hammett(1993)]{Dorland1993GyrofluidEffects}
{\sc \au{Dorland, W} \& \au{Hammett, G~W}} \yr{1993}  \at{{Gyrofluid turbulence models with kinetic effects}}.  \jt{Physics of Fluids B: Plasma Physics}  \bvol{5}~(3),  \pg{812--835}.

\bibitem[Dorland {\em et~al.\/}(2000)Dorland, Jenko, Kotschenreuther \& Rogers]{Dorland2000ElectronTurbulence}
{\sc \au{Dorland, W.}, \au{Jenko, F.}, \au{Kotschenreuther, M.} \& \au{Rogers, B.~N.}} \yr{2000}  \at{{Electron Temperature Gradient Turbulence}}.  \jt{Physical Review Letters}  \bvol{85}~(26),  \pg{5579--5582}.

\bibitem[Dougherty(1964)]{Dougherty1964ModelSolution}
{\sc \au{Dougherty, J~P}} \yr{1964}  \at{{Model Fokker-Planck Equation for a Plasma and Its Solution}}.  \jt{The Physics of Fluids}  \bvol{7}~(11),  \pg{1788--1799}.

\bibitem[Dudt {\em et~al.\/}(2023)Dudt, Conlin, Panici \& Kolemen]{Dudt2023TheOptimization}
{\sc \au{Dudt, D.W.}, \au{Conlin, R.}, \au{Panici, D.} \& \au{Kolemen, E.}} \yr{2023}  \at{{The DESC stellarator code suite Part 3: Quasi-symmetry optimization}}.  \jt{Journal of Plasma Physics}  \bvol{89}~(2),  \pg{955890201}.

\bibitem[Dudt \& Kolemen(2020)]{Dudt2020DESC:Solver}
{\sc \au{Dudt, D.~W.} \& \au{Kolemen, E.}} \yr{2020}  \at{{DESC: A stellarator equilibrium solver}}.  \jt{Physics of Plasmas}  \bvol{27}~(10),  \pg{102513}.

\bibitem[Faber {\em et~al.\/}(2015)Faber, Pueschel, Proll, Xanthopoulos, Terry, Hegna, Weir, Likin \& Talmadge]{Faber2015GyrokineticStellarator}
{\sc \au{Faber, B~J}, \au{Pueschel, M~J}, \au{Proll, J H~E}, \au{Xanthopoulos, P}, \au{Terry, P~W}, \au{Hegna, C~C}, \au{Weir, G~M}, \au{Likin, K~M} \& \au{Talmadge, J~N}} \yr{2015}  \at{{Gyrokinetic studies of trapped electron mode turbulence in the Helically Symmetric eXperiment stellarator}}.  \jt{Physics of Plasmas}  \bvol{22}~(7),  \pg{072305}.

\bibitem[Frieman \& Chen(1982)]{Frieman1982NonlinearEquilibria}
{\sc \au{Frieman, E~A} \& \au{Chen, Liu}} \yr{1982}  \at{{Nonlinear gyrokinetic equations for low‐frequency electromagnetic waves in general plasma equilibria}}.  \jt{The Physics of Fluids}  \bvol{25}~(3),  \pg{502--508}.

\bibitem[Gonz{\'{a}}lez-Jerez {\em et~al.\/}(2022{\natexlab{{\em a\/}}})Gonz{\'{a}}lez-Jerez, Xanthopoulos, Garc{\'{i}}a-Rega{\~{n}}a, Calvo, Alcus{\'{o}}n, Ba{\~{n}}{\'{o}}n~Navarro, Barnes, Parra \& Geiger]{Gonzalez-Jerez2022ElectrostaticGENEb}
{\sc \au{Gonz{\'{a}}lez-Jerez, A.}, \au{Xanthopoulos, P.}, \au{Garc{\'{i}}a-Rega{\~{n}}a, J.M.}, \au{Calvo, I.}, \au{Alcus{\'{o}}n, J.}, \au{Ba{\~{n}}{\'{o}}n~Navarro, A.}, \au{Barnes, M.}, \au{Parra, F.I.} \& \au{Geiger, J.}} \yr{2022{\natexlab{{\em a\/}}}}  \at{{Electrostatic gyrokinetic simulations in Wendelstein 7-X geometry: benchmark between the codes stella and GENE}}.  \jt{Journal of Plasma Physics}  \bvol{88}~(3),  \pg{905880310}.

\bibitem[Gonz{\'{a}}lez-Jerez {\em et~al.\/}(2022{\natexlab{{\em b\/}}})Gonz{\'{a}}lez-Jerez, Xanthopoulos, Garc{\'{i}}a-Rega{\~{n}}a, Calvo, Alcus{\'{o}}n, Ba{\~{n}}{\'{o}}n~Navarro, Barnes, Parra \& Geiger]{Gonzalez-Jerez2022ElectrostaticGENE}
{\sc \au{Gonz{\'{a}}lez-Jerez, A.}, \au{Xanthopoulos, P.}, \au{Garc{\'{i}}a-Rega{\~{n}}a, J.M.}, \au{Calvo, I.}, \au{Alcus{\'{o}}n, J.}, \au{Ba{\~{n}}{\'{o}}n~Navarro, A.}, \au{Barnes, M.}, \au{Parra, F.I.} \& \au{Geiger, J.}} \yr{2022{\natexlab{{\em b\/}}}}  \at{{Electrostatic gyrokinetic simulations in Wendelstein 7-X geometry: benchmark between the codes stella and GENE}}.  \jt{Journal of Plasma Physics}  \bvol{88}~(3),  \pg{905880310}.

\bibitem[Goodman {\em et~al.\/}(2022)Goodman, Mata, Henneberg, Jorge, Landreman, Plunk, Smith, Mackenbach \& Helander]{Goodman2022ConstructingFields}
{\sc \au{Goodman, Alan}, \au{Mata, Katia~Camacho}, \au{Henneberg, Sophia~A}, \au{Jorge, Rogerio}, \au{Landreman, Matt}, \au{Plunk, Gabriel}, \au{Smith, Hakan}, \au{Mackenbach, Ralf} \& \au{Helander, Per}} \yr{2022}  \at{{Constructing precisely quasi-isodynamic magnetic fields}} .

\bibitem[Greene \& Chance(1981)]{Greene1981TheModes}
{\sc \au{Greene, J.M.} \& \au{Chance, M.S.}} \yr{1981}  \at{{The second region of stability against ballooning modes}}.  \jt{Nuclear Fusion}  \bvol{21}~(4),  \pg{453--464}.

\bibitem[Hegna {\em et~al.\/}(2018)Hegna, Terry \& Faber]{Hegna2018TheoryTransport}
{\sc \au{Hegna, C~C}, \au{Terry, P~W} \& \au{Faber, B~J}} \yr{2018}  \at{{Theory of ITG turbulent saturation in stellarators: Identifying mechanisms to reduce turbulent transport}}.  \jt{Physics of Plasmas}  \bvol{25}~(2),  \pg{022511}.

\bibitem[Helander(2014)]{Helander2014TheoryFields}
{\sc \au{Helander, P.}} \yr{2014}  \at{{Theory of plasma confinement in non-axisymmetric magnetic fields}}.  \jt{Reports on Progress in Physics}  \bvol{77}~(8),  \pg{087001}.

\bibitem[Helander {\em et~al.\/}(2013)Helander, Proll \& Plunk]{Helander2013CollisionlessModes}
{\sc \au{Helander, P}, \au{Proll, J H~E} \& \au{Plunk, G~G}} \yr{2013}  \at{{Collisionless microinstabilities in stellarators. I. Analytical theory of trapped-particle modes}}.  \jt{Physics of Plasmas}  \bvol{20}~(12),  \pg{122505}.

\bibitem[Horton(1999)]{Horton1999DriftTransport}
{\sc \au{Horton, W.}} \yr{1999}  \at{{Drift waves and transport}}.  \jt{Reviews of Modern Physics}  \bvol{71}~(3),  \pg{735--778}.

\bibitem[Jorge {\em et~al.\/}(2023)Jorge, Dorland, Kim, Landreman, Mandell, Merlo \& Qian]{Jorge2023DirectDevices}
{\sc \au{Jorge, R.}, \au{Dorland, W.}, \au{Kim, P.}, \au{Landreman, M.}, \au{Mandell, N.~R.}, \au{Merlo, G.} \& \au{Qian, T.}} \yr{2023}  \at{{Direct Microstability Optimization of Stellarator Devices}} .

\bibitem[Kotschenreuther {\em et~al.\/}(1995{\natexlab{{\em a\/}}})Kotschenreuther, Dorland, Beer \& Hammett]{Kotschenreuther1995QuantitativeEffects}
{\sc \au{Kotschenreuther, M}, \au{Dorland, W}, \au{Beer, M~A} \& \au{Hammett, G~W}} \yr{1995{\natexlab{{\em a\/}}}}  \at{{Quantitative predictions of tokamak energy confinement from first‐principles simulations with kinetic effects}}.  \jt{Physics of Plasmas}  \bvol{2}~(6),  \pg{2381--2389}.

\bibitem[Kotschenreuther {\em et~al.\/}(1995{\natexlab{{\em b\/}}})Kotschenreuther, Rewoldt \& Tang]{Kotschenreuther1995ComparisonInstabilities}
{\sc \au{Kotschenreuther, Mike}, \au{Rewoldt, G.} \& \au{Tang, W.~M.}} \yr{1995{\natexlab{{\em b\/}}}}  \at{{Comparison of initial value and eigenvalue codes for kinetic toroidal plasma instabilities}}.  \jt{Computer Physics Communications}  \bvol{88}~(2-3),  \pg{128--140}.

\bibitem[Landreman {\em et~al.\/}(2022)Landreman, Buller \& Drevlak]{Landreman2022OptimizationConfinement}
{\sc \au{Landreman, M}, \au{Buller, S} \& \au{Drevlak, M}} \yr{2022}  \at{{Optimization of quasi-symmetric stellarators with self-consistent bootstrap current and energetic particle confinement}}.  \jt{Physics of Plasmas}  \bvol{29}~(8),  \pg{082501}.

\bibitem[Landreman \& Paul(2022)]{Landreman2022MagneticConfinement}
{\sc \au{Landreman, Matt} \& \au{Paul, Elizabeth}} \yr{2022}  \at{{Magnetic Fields with Precise Quasisymmetry for Plasma Confinement}}.  \jt{Physical Review Letters}  \bvol{128}~(3),  \pg{035001}.

\bibitem[Mandell {\em et~al.\/}(2022)Mandell, Dorland, Abel, Gaur, Kim, Martin \& Qian]{Mandell2022GX:Design}
{\sc \au{Mandell, N.~R.}, \au{Dorland, W.}, \au{Abel, I.}, \au{Gaur, R.}, \au{Kim, P.}, \au{Martin, M.} \& \au{Qian, T.}} \yr{2022}  \at{{GX: a GPU-native gyrokinetic turbulence code for tokamak and stellarator design}} .

\bibitem[Mandell {\em et~al.\/}(2018)Mandell, Dorland \& Landreman]{Mandell2018Laguerre-HermiteGyrokinetics}
{\sc \au{Mandell, N~R}, \au{Dorland, W} \& \au{Landreman, M}} \yr{2018}  \at{{Laguerre-Hermite pseudo-spectral velocity formulation of gyrokinetics}}.  \jt{J. Plasma Phys}  \bvol{84}~(1),  \pg{905840108}.

\bibitem[Mariani {\em et~al.\/}(2018)Mariani, Brunner, Dominski, Merle, Merlo, Sauter, G{\"{o}}rler, Jenko \& Told]{Mariani2018IdentifyingFluxes}
{\sc \au{Mariani, A}, \au{Brunner, S}, \au{Dominski, J}, \au{Merle, A}, \au{Merlo, G}, \au{Sauter, O}, \au{G{\"{o}}rler, T}, \au{Jenko, F} \& \au{Told, D}} \yr{2018}  \at{{Identifying microturbulence regimes in a TCV discharge making use of physical constraints on particle and heat fluxes}}.  \jt{Physics of Plasmas}  \bvol{25}~(1),  \pg{012313}.

\bibitem[McKinney {\em et~al.\/}(2019)McKinney, Pueschel, Faber, Hegna, Talmadge, Anderson, Mynick \& Xanthopoulos]{McKinney2019AQuasi-axisymmetricstellarator}
{\sc \au{McKinney, I.~J.}, \au{Pueschel, M.~J.}, \au{Faber, B.~J.}, \au{Hegna, C.~C.}, \au{Talmadge, J.~N.}, \au{Anderson, D.~T.}, \au{Mynick, H.~E.} \& \au{Xanthopoulos, P.}} \yr{2019}  \at{{A comparison of turbulent transport in a quasi-helical and a quasi-axisymmetric stellarator}}.  \jt{Journal of Plasma Physics}  \bvol{85}~(5),  \pg{905850503}.

\bibitem[Mynick {\em et~al.\/}(2010)Mynick, Pomphrey \& Xanthopoulos]{Mynick2010OptimizingTransport}
{\sc \au{Mynick, H.~E.}, \au{Pomphrey, N.} \& \au{Xanthopoulos, P.}} \yr{2010}  \at{{Optimizing Stellarators for Turbulent Transport}}.  \jt{Physical Review Letters}  \bvol{105}~(9),  \pg{095004}.

\bibitem[Panici {\em et~al.\/}(2023)Panici, Conlin, Dudt, Unalmis \& Kolemen]{Panici2023TheComputations}
{\sc \au{Panici, D.}, \au{Conlin, R.}, \au{Dudt, D.W.}, \au{Unalmis, K.} \& \au{Kolemen, E.}} \yr{2023}  \at{{The DESC stellarator code suite. Part 1. Quick and accurate equilibria computations}}.  \jt{Journal of Plasma Physics}  \bvol{89}~(3),  \pg{955890303}.

\bibitem[Proll {\em et~al.\/}(2022)Proll, Plunk, Faber, G{\"{o}}rler, Helander, McKinney, Pueschel, Smith \& Xanthopoulos]{Proll2022TurbulenceGradient}
{\sc \au{Proll, J.H.E.}, \au{Plunk, G.G.}, \au{Faber, B.J.}, \au{G{\"{o}}rler, T.}, \au{Helander, P.}, \au{McKinney, I.J.}, \au{Pueschel, M.J.}, \au{Smith, H.M.} \& \au{Xanthopoulos, P.}} \yr{2022}  \at{{Turbulence mitigation in maximum-J stellarators with electron-density gradient}}.  \jt{Journal of Plasma Physics}  \bvol{88}~(1),  \pg{905880112}.

\bibitem[Proll {\em et~al.\/}(2012)Proll, Helander, Connor \& Plunk]{Proll2012ResilienceInstabilities}
{\sc \au{Proll, J. H.~E.}, \au{Helander, P.}, \au{Connor, J.~W.} \& \au{Plunk, G.~G.}} \yr{2012}  \at{{Resilience of Quasi-Isodynamic Stellarators against Trapped-Particle Instabilities}}.  \jt{Physical Review Letters}  \bvol{108}~(24),  \pg{245002}.

\bibitem[Proll {\em et~al.\/}(2016)Proll, Mynick, Xanthopoulos, Lazerson \& Faber]{Proll2016TEMStellarators}
{\sc \au{Proll, J H~E}, \au{Mynick, H~E}, \au{Xanthopoulos, P}, \au{Lazerson, S~A} \& \au{Faber, B~J}} \yr{2016}  \at{{TEM turbulence optimisation in stellarators}}.  \jt{Plasma Physics and Controlled Fusion}  \bvol{58}~(1),  \pg{014006}.

\bibitem[Qian {\em et~al.\/}(2022)Qian, Buck, Gaur, Mandell, Kim \& Dorland]{Qian2022StellaratorGX}
{\sc \au{Qian, Tony}, \au{Buck, Braden}, \au{Gaur, Rahul}, \au{Mandell, Noah}, \au{Kim, Patrick} \& \au{Dorland, William}} \yr{2022} {Stellarator profile predictions using Trinity3D and GX}.  \bt{In {\em Bull. Am. Phys. Soc.\/}}.  \publ{Spokane: American Physical Society}.

\bibitem[Roberg-Clark {\em et~al.\/}(2022)Roberg-Clark, Xanthopoulos \& Plunk]{Roberg-Clark2022ReductionOptimization}
{\sc \au{Roberg-Clark, G.~T.}, \au{Xanthopoulos, P.} \& \au{Plunk, G.~G.}} \yr{2022}  \at{{Reduction of electrostatic turbulence in a quasi-helically symmetric stellarator via critical gradient optimization}} .

\bibitem[S{\'{a}}nchez {\em et~al.\/}(2023)S{\'{a}}nchez, Velasco, Calvo \& Mulas]{Sanchez2023A}
{\sc \au{S{\'{a}}nchez, E.}, \au{Velasco, J.L.}, \au{Calvo, I.} \& \au{Mulas, S.}} \yr{2023}  \at{{A quasi-isodynamic configuration with good confinement of fast ions at low plasma {$\beta$}}}.  \jt{Nuclear Fusion}  \bvol{63}~(6),  \pg{066037}.

\bibitem[Spall(1987)]{Spall1987AEstimates}
{\sc \au{Spall, James~C}} \yr{1987} {A Stochastic Approximation Technique for Generating Maximum Likelihood Parameter Estimates}.  \bt{In {\em 1987 American Control Conference\/}},  \pg{pp. 1161--1167}.

\bibitem[Velasco {\em et~al.\/}(2023)Velasco, Calvo, S{\'{a}}nchez \& Parra]{Velasco2023RobustFields}
{\sc \au{Velasco, J.L.}, \au{Calvo, I.}, \au{S{\'{a}}nchez, E.} \& \au{Parra, F.I.}} \yr{2023}  \at{{Robust stellarator optimization via flat mirror magnetic fields}}.  \jt{Nuclear Fusion}  \bvol{63}~(12),  \pg{126038}.

\end{thebibliography}

\end{document}